**Automated Diagnosis of Cardiovascular Diseases from Cardiac Magnetic Resonance Imaging Using Deep Learning Models: A Review**

Mahboobeh Jafari[1], Afshin Shoeibi[1,2,*], Marjane Khodatars[3], Navid Ghassemi[1], Parisa Moridian[1], Niloufar Delfan[4], Roohallah Alizadehsani[5], Abbas Khosravi[5], Sai Ho Ling[6], Yu-Dong Zhang[7], Shui-Hua Wang[7], Juan M. Gorriz[2,8], Hamid Alinejad Rokny[9,10,11], U. Rajendra Acharya[12,13,14]

[1] Internship in BioMedical Machine Learning Lab, The Graduate School of Biomedical Engineering, UNSW Sydney, Sydney, NSW, 2052, Australia.

[2] Data Science and Computational Intelligence Institute, University of Granada, Spain.

[3] Department of Medical Engineering, Mashhad Branch, Islamic Azad University, Mashhad, Iran.

[4] Faculty of Computer Engineering, Dept. of Artificial Intelligence Engineering, K. N. Toosi University of Technology, Tehran, Iran.

[5] Intelligent for Systems Research and Innovation (IISRI), Deakin University, Victoria 3217, Australia.

[6] Faculty of Engineering and IT, University of Technology Sydney (UTS), Australia.

[7] School of Computing and Mathematical Sciences, University of Leicester, Leicester, UK.

[8] Department of Psychiatry, University of Cambridge, UK.

[9] BioMedical Machine Learning Lab, The Graduate School of Biomedical Engineering, UNSW Sydney, Sydney, NSW, 2052, Australia.

[10] UNSW Data Science Hub, The University of New South Wales, Sydney, NSW, 2052, Australia.

[11] Health Data Analytics Program, AI-enabled Processes (AIP) Research Centre, Macquarie University, Sydney, 2109, Australia.

[12] Ngee Ann Polytechnic, Singapore 599489, Singapore.

[13] Dept. of Biomedical Informatics and Medical Engineering, Asia University, Taichung, Taiwan.

[14] Dept. of Biomedical Engineering, School of Science and Technology, Singapore University of Social Sciences, Singapore.

* Corresponding author: Afshin Shoeibi (Afshin.shoeibi@gmail.com)

## Abstract

In recent years, cardiovascular diseases (CVDs) have become one of the leading causes of mortality globally. CVDs appear with minor symptoms and progressively get worse. The majority of people experience symptoms such as exhaustion, shortness of breath, ankle swelling, fluid retention, and other symptoms when starting CVD. Coronary artery disease (CAD), arrhythmia, cardiomyopathy, congenital heart defect (CHD), mitral regurgitation, and angina are the most common CVDs. Clinical methods such as blood tests, electrocardiography (ECG) signals, and medical imaging are the most effective methods used for the detection of CVDs. Among the diagnostic methods, cardiac magnetic resonance imaging (CMR) is increasingly used to diagnose, monitor the disease, plan treatment and predict CVDs. Coupled with all the advantages of CMR data, CVDs diagnosis is challenging for physicians due to many slices of data, low contrast, etc. To address these issues, deep learning (DL) techniques have been employed to the diagnosis of CVDs using CMR data, and much research is currently being conducted in this field. This review provides an overview of the studies performed in CVDs detection using CMR images and DL techniques. The introduction section examined CVDs types, diagnostic methods, and the most important medical imaging techniques. In the following, investigations to detect CVDs using CMR images and the most significant DL methods are presented. Another section discussed the challenges in diagnosing CVDs from CMR data. Next, the discussion section discusses the results of this review, and future work in CVDs diagnosis from CMR images and DL techniques are outlined. The most important findings of this study are presented in the conclusion section.

## 1. Introduction

CVDs are one of the most common causes of death and endanger the health of many people around the world annually [1-2]. According to the World Health Organization (WHO), CVDs are the leading cause of human death worldwide [3-4]. According to this statistics, 17.9 million people died from CVDs in 2016, accounting for 31% of all global deaths [5-7]. In addition, coronary heart disease and stroke are responsible for four out of five deaths from CVDs, and one-third of these deaths occur in people under 70 years [8-10]. Some of the most important CVDs include coronary arteries disease (CAD) [11-12], rheumatoid arthritis [13-14], myocarditis [15-17], cardiovascular diabetes [18-19], etc. Figure (1) shows the patients with cardiovascular diabetes in the world.

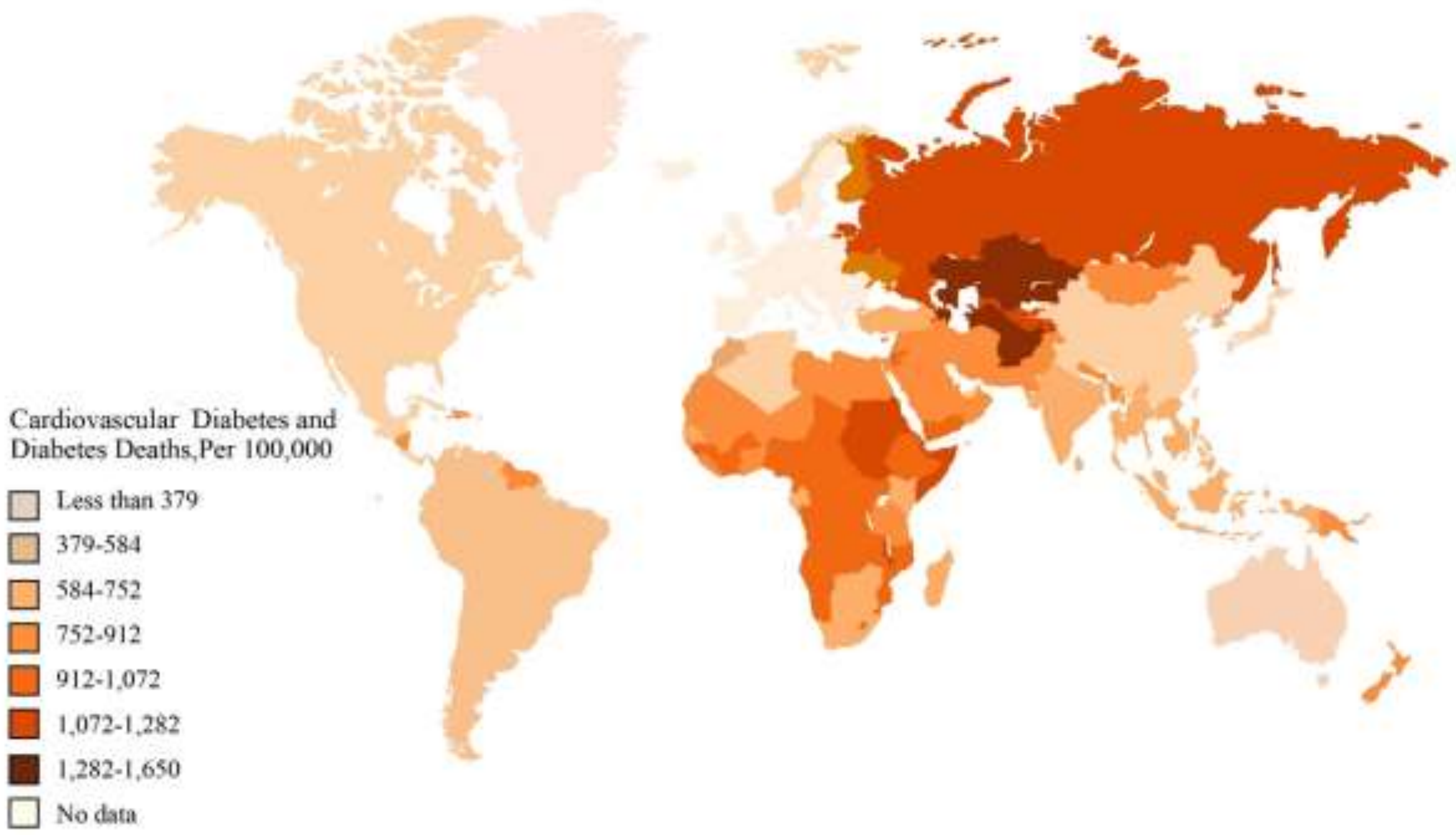

Fig. 1. Patients with cardiovascular diabetes in the world.

The human heart is responsible for pumping blood and circulating it throughout the body [18], so any abnormality in it results in CVDs [19]. CAD is considered the most common type of CVD [20-22]. CAD is the plaque accumulation in the arteries that supply oxygen-rich blood to the heart [20-22]. Plaque causes narrowing or blockage, which restricts blood flow and thus reduces blood oxygen to parts of the heart [20-22]. Some of the most significant symptoms of CAD involve chest pain or discomfort and shortness of breath [20-22]. Cardiac arrhythmia is another of the most prevalent CVDs caused by atrial fibrillation and ventricular arrhythmias [23-24]. A cardiac arrhythmia occurs due to a non-uniform heartbeat. Weakness and pain in the chest area are the most important symptoms of arrhythmia [23-24]. Congenital heart disease (CHD) is another human CVDs.  There is a defect in the structure of the heart or large arteries are present by birth [25-26]. Signs and symptoms of CHD include rapid breathing, a blue tingein the skin (cyanosis), poor weight gain, and tiredness [25-26]. Figure (2) shows the common CVDs together with their details. In recent years, major advances in cardiac research have been made to improve the diagnosis and treatment of CVDs as well as decline their case fatality.

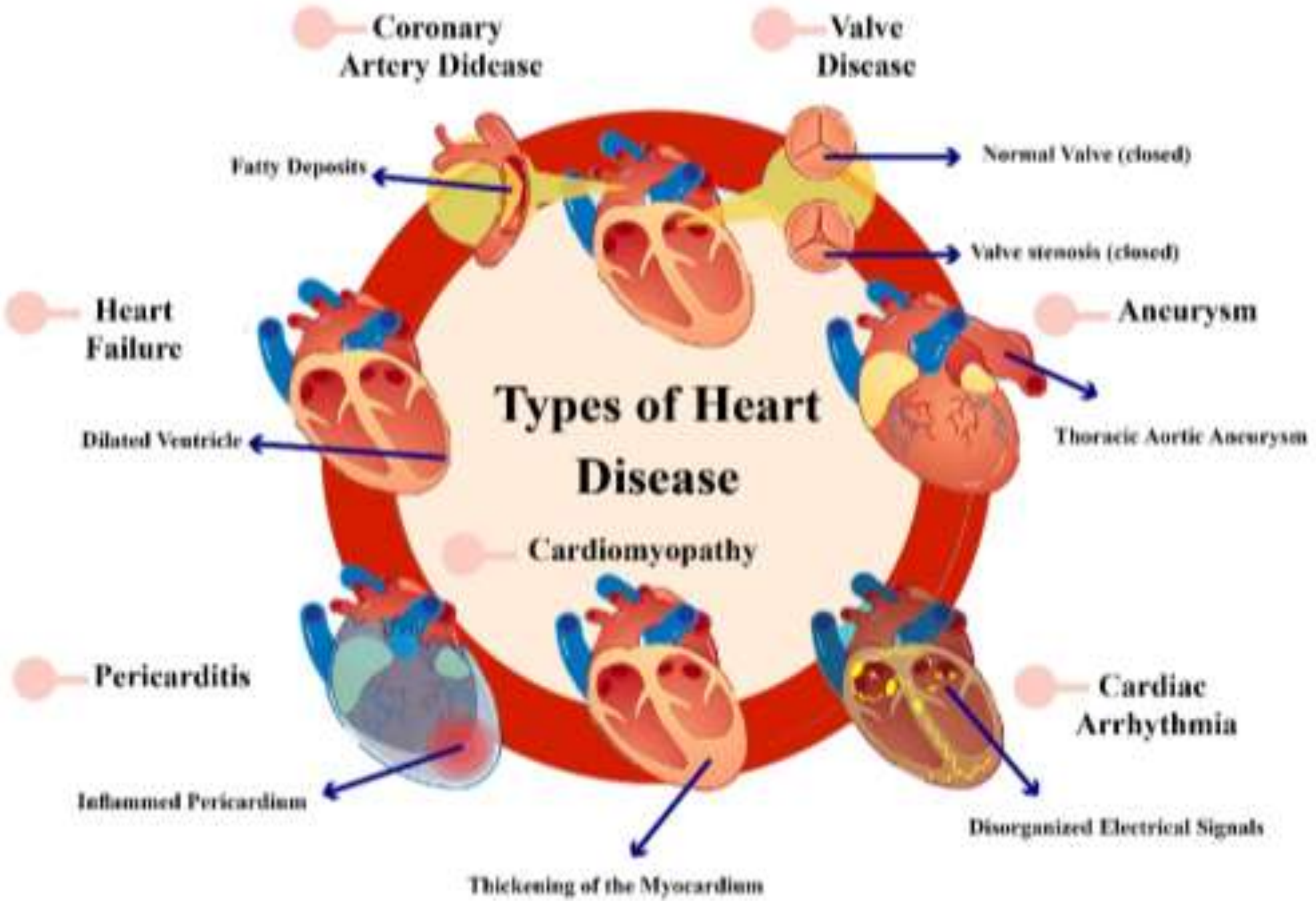

Fig. 2. Common types of CVDs with details.

Cardiac ultrasound or echocardiography (Echo) works by utilizing sound waves which is a non-invasive modality to image heart tissue [27-28]. In this method, ultrasound waves are taken advantage of to produce echocardiography images of the heart [27-28]. Echo helps physicians detect various types of CVDs by assessing the heart's structure, analyzing how the blood flows in them, and evaluating the heart's pumping cavities [27-28]. Advantages of echocardiography include readily accessible, portability, high temporal resolution, and no ionizing radiation [29].

CT is a non-invasive imaging technique that can be applied to detect a variety of CVDs, brain diseases, etc. [30-31]. In particular, cardiac CT provides the anatomical evaluation of the heart, especially CAD [32]. This imaging technique involves two techniques: non-contrast CT and contrast-enhanced coronary CT angiography (CTA) [30-31]. Non-contrast CT makes use of the density of tissues to generate the image so that various densities can be simply distinguished using different attenuation values [30-31] [33-34]. In addition, the amount of calcium in the coronary arteries can be calculated using non-contrast CT [30-31] [33-34]. In comparison, contrast-enhanced coronary CTA provides the ability to generate extraordinary images of the heart, arteries, and coronary arteries [30-31] [33-34]. Radiation exposure is one of the major weaknesses of cardiac CT imaging. Frequent exposure to radiation is associated with deleterious health effects, including an increased cancer risk [30-31] [33-34].

CMR imaging offers an excellent quantitative assessment of cardiac chamber volume/function [374] and the extent of myocardial infarction/fibrosis [375]. It is a guideline-recommended modality for the diagnosis of diverse CVDs, including ischemic heart disease [37, 38], heritable or acquired cardiomyopathy [39], myocarditis [40], congenital heart disease [41], etc. For measurement of ventricular volume, function, and mass, accurate segmentation of the endocardial (and, in the case of myocardial mass, epicardial) contours on standard cine CMR images is a necessary prerequisite. Typically, the contours are drawn —either manually or software-assisted—on a stack of contiguous parallel slices of two-dimensional (2D) short-axis time-series cine CMR images of the ventricles at desired phases of the cardiac cycle, e.g., end-diastole and -systole, to derive the corresponding time-aligned three-dimensional (3D) ventricular volumes using Simpson's method of disc without the need for geometric assumption [376]. Indeed, cine CMR analysis is the gold standard for right ventricular (RV) volume/function measurement as the RV can be optimally visualized on CMR without being limited by issues of acoustic window access, as with echocardiography [377]. Late gadolinium enhancement (LGE) [378] is an established CMR imaging technique in which images acquired ten to twenty minutes after gadolinium-based contrast administration are used to define in granular detail regions of myocardial infarct, fibrosis, infiltrate, etc. Indeed, segmentation can also be performed to outline and quantitate areas of abnormal tissue, e.g., myocardial infarct [379], microvascular obstruction [380], and non-infarct fibrosis [381], which may have prognostic significance.

In addition to quantitative measurements, CMR must be qualitatively interpreted by medical experts, which is time-consuming and subject to human bias. The presence of noise and imaging artifacts can further confound the interpretation, potentially resulting in misdiagnosis. However, CMRI data is the gold standard and most popular procedure for diagnosing cardiac diseases among physicians. To address CMRI challenges, researchers have proposed artificial intelligence (AI) techniques for the automatic diagnosis of CVDs using CMRI data [1-10]. In the presented papers, the main objective of the researchers is to achieve a tool for rapid detection of CVDs using CMRI images along with AI techniques. For this purpose, the researchers have conducted extensive research on ML-based approaches for diagnosing CVDs from CMRI data, including introducing various segmentation and classification approaches [42-44]. However, ML methods presented satisfactory results in early research on the diagnosis of CVDs. Nevertheless, due to high computational complexity, and inefficient performance with huge databases these methods were not able yield good performances. To tackle the challenges of ML methods, AI researchers introduced DL methods [45-47]. DL networks were able to overcome the limitations of ML methods [45-47]. The DL models were employed in various medical applications, including the diagnosis of CVDs [4], and reported satisfactory results. Researchers hope that in the near future, an accurate software platform for diagnosing CVDs using MRI data and DL techniques will be realized.

In this study, papers in diagnosis of CVDs using CMRI images and DL techniques were examined. The section 3 describes search strategy papers regarding preferred reporting items for systematic reviews and meta-analyses (PRISMA) guidelines [48]. In section 4, the conducted review papers in diagnosis of CVDs are studied. The computer aided diagnosis system (CADS) and their steps for diagnosis CVDs from CMRI images are provided in Section 5. This section discusses in datasets, preprocessing, and popular DL models for diagnosis of CVDs. Also, in this section, segmentation, classification, and fusion research based on DL methods are summarized in different Tables. Section 6 is allocated to the most important challenges in diagnosis of CVDs using CMRI data. The discussion of this paper, along with its details, is provided in section 7. Future work is also presented in section that suggest potential directions for future works. Finally, the conclusion and the findings of this study are discussed in the section 7.

## 2. Search strategy

This section searches papers based on PRISMA guidelines [48]. We have searched the papers published between 2016 and 2022 in the field of heart diseases using the general keywords "CVDs", "deep learning", "Segmentation", "classification", and "CMRI". Keyword searches are performed in repositories such as Science Direct, Frontiers, MDPI, IEEE Xplore, Nature, Springer, ArXiv, and Wiley citation databases.

The selection method of important articles for diagnosing CVDs with AI techniquesis presented in this section. The selection process of papers related to this field has been done in three levels. In Figure (3), the review process of papers based on PRISMA guidelines is provided. First, 324 articles were collected and then 38 articles were filtered out as they are not related to this area of research. In the following, 33 papers did not use the CMRI datasets and filtered. Finally, 21 articles were filtered out as they did not use DL techniques in their studies. Figure (3) displayed the PRISMA guidelines for diagnosis of CVDs from CMR images using DL methods. In addition, the exclusion and inclusion criteria used in this work are provided in Table (1).

Table 1. Exclusion and inclusion criteria used for the diagnosis of CVDs.

| Inclusion | Exclusion |
|---|---|
| 1. CMRI Images<br>3. Different types of CVDs.<br>3. CVDs detection<br><br>4. DL models (CNNs, RNNs, AEs, CNN-RNN, CNN-AE, GAN, Transfer Learning, etc.) | 1. Treatment of CVDs<br>2. Clinical methods for CVDs treatment<br>3. Rehabilitation systems for CVDs detection (Without AI techniques) |

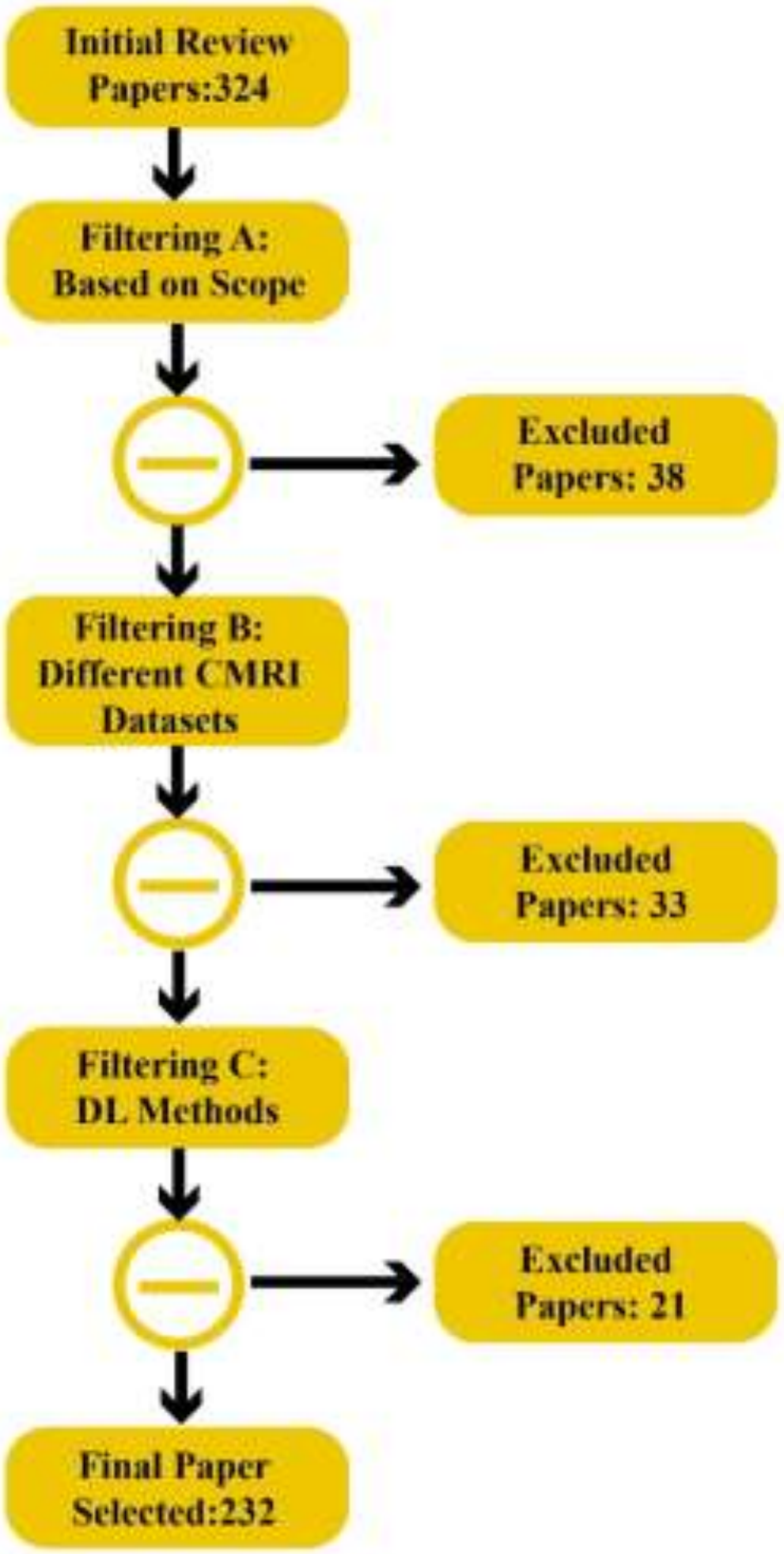

Fig. 3. Literature search procedure.

## 3. Review studies on AI-enabled image segmentation

Leiner et al. [11] reviewed advances in ML for image reconstruction, feature extraction, image analysis, and diagnostic evaluation of CMR images. They also highlighted important areas of research like image reconstruction, improving spatial and temporal resolution, perfusion analysis, and myocardial mapping.

Segmentation of CMR images is an important area used for quantitative CMR assessment, including calculation of heart chamber volumes (and function) and delineation of abnormal myocardial tissues (e.g., myocardial infarct, fibrosis, etc.). Several review papers have been published on the use of AI for segmenting cardiac and vascular structures as well as tissues like fat and scars on different imaging modalities such as CMR, computed tomography, echocardiography, etc. [2] [6] [9]. Litjens et al. [9] reviewed 80 papers on the diagnosis of CVD using CMR and other modalities. In this review paper, we discussed the most important DL models, including convolutional neural networks (CNNs). In addition, we reviewed few novel DL models, such as generative adversarial networks (GANs).

Studies on automated segmentation of the left ventricle (LV) on short-axis cine CMR images using ML [14] and DL [1] architectures have been reviewed in [14] and [1], respectively. In [10], 3D convolution architectures for handling volumetric LV CMR datasets were reviewed. Compared with 2D networks, 3D can capture the entirety of the spatial information while reducing the number of training data. Still, the high memory requirement limits the network depth and the filter's field of view. Wu et al. [8] also reviewed the papers on the segmentation of fibrosis and scars on late gadolinium enhancement (LGE) CMR images using DL models. In [4-5], AI methods for segmenting and quantitating regions of atrial fibrosis on LGE CMR images—which have diagnostic and prognostic implications in conditions like atrial fibrillation—were reviewed, and the challenges were discussed. The review papers published on image segmentation using AI methods are summarized in Table (2).

Table 2. Summary of review papers on image segmentation using AI.

| Work | Year | Image input | Segmentation | Methods |
|---|---|---|---|---|
| [6] | 2021 | Multi-modal | Chamber and vessel borders, tissue | ML, DL |
| [2] | 2020 | Multi-modal | Chamber and vessel borders, fibrosis | DL |
| [9] | 2019 | Multi-modal | Chamber and vessel borders, fibrosis | DL |
| [14] | 2018 | Cine CMR | LV myocardial border | ML |
| [13] | 2022 | Cine CMR | LV myocardial border | ML, DL |
| [10] | 2020 | 3D cine CMR | LV myocardial border | DL |
| [3] | 2019 | Cine CMR | LV and RV myocardial borders | ML, DL |
| [1] | 2021 | Cine CMR | RV myocardial border | ML, DL |
| [8] | 2021 | LGE-CMR | LV and LA fibrosis | DL |
| [4] | 2020 | LGE-CMR | LA fibrosis | DL |
| [5] | 2022 | LGE-CMR | LA fibrosis | ML, DL |

## 4. Computer-Aided Diagnosis for Heart disease

Early detection of CVDs from CMR images extends patients' lifespan and quality of life. As mentioned in Section 3, many papers have been published for diagnosing CVDs using CMR images and AI methods. The main aim of research in CADS based on AI methods is to assist clinicians in interpreting CMR data for early detection of CVDs [1]. In general, researchers exploit ML and DL techniques in the implementation of CADS to diagnose CVDs [1-19]. DL models are state-of-the-art AI techniques that are evolving rapidly. For this purpose, the application of DL techniques in diagnosing CVDs has grown dramatically in recent years [8] [10]. The main objective of this study is to enhance the performance of CADS to assist doctors in the accurate diagnosis of CVDs. The CADS based on DL networks for the diagnosis of CVDs consists of CMR datasets, preprocessing, DL models, and evaluation parameters steps. Figure (4) illustrates the block diagram for CVDs detection using DL methods. To achieve better performance using CADS, CMR images are pre-processed to remove artifacts, increase contrast, etc. In the next step, DL models are fed with CMR images for diagnosis of CVDs. Ultimately, the evaluation criteria demonstrate the effectiveness of the proposed CADS for CVDs detection. As aforementioned, the diagnosis of CVDs is largely reliable on the physician's subjective interpretation. CADS based on DL can alleviate this subjectivity by improving the detection of CVDs and quantitative support for decision-making [14]. The following details of each CADS section for CVDs detection based on DL models are provided.

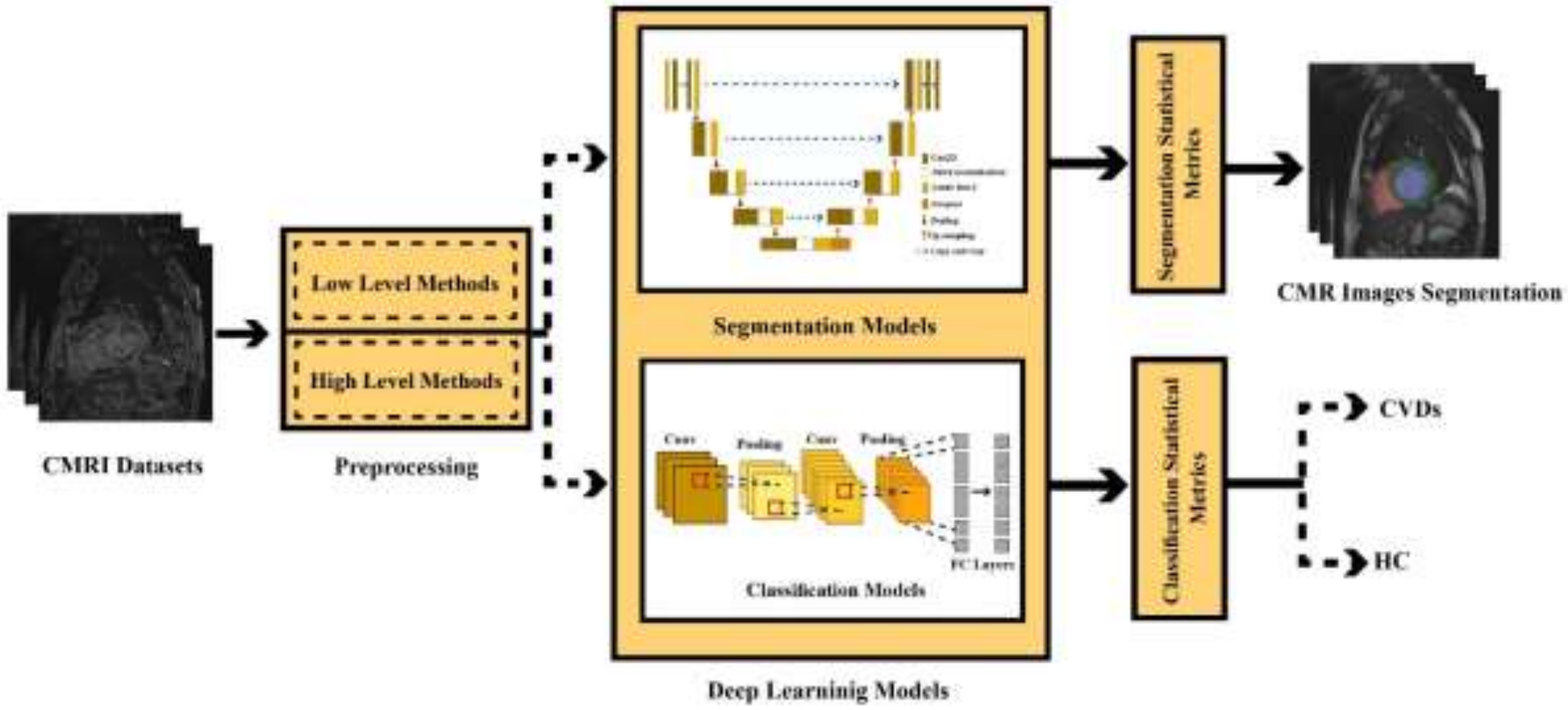

Fig. 4. Illustration of CVDs detection using DL methods.

### 4.1. Datasets

Datasets play an important role in the DL-based diagnosis of CVDs. To date, several datasets have been made available to researchers for CVD diagnosis, including ECG, echocardiography, and CMR. This section presents the most important available CMR datasets for CVDs detection. The remainder of this section describes the available datasets from CMR data. Also, a summary of CMR datasets available is summarized in Table (3).

#### 4.1.1. Sunnybrook Cardiac Data (SCD)

The SCD dataset contains cine-CMR images of 45 individuals with four pathologies, namely, healthy, hypertrophy, heart failure with infarction and heart failure without infarction, and with medical interpretations [49]. In this dataset, images are acquired while the patient holds his breath for 10-15 seconds with a time resolution of 20 cardiac phases in the cardiac cycle [49]. The data is in DICOM format and includes Metadata parameters about the patient and the image. A set of contours is delineated manually for each patient record at end-diastolic (ED) and end-systolic (ES) slices [49]. These contours were drawn by Perry Radau of the Sunnybrook Health Science Center [49]. The data is provided for analysis by physicians without any pre-processing. In SCD, data is randomly split into three groups: 15 for training, 15 for testing, and 15 for an online challenge [49]. The training dataset contains CMR images and their ground truth for segmentation application. Furthermore, the test dataset does not involve ground truth for segmentation [49]. In 2009, a subset of the SCD dataset was first utilized in the myocardial segmentation challenge with CMR, held by a MICCAI workshop. The entire dataset is now available in the CAP [49].

#### 4.1.2. The Automated Cardiac Diagnosis Challenge (ACDC) MICCAI 2017 Challenge

The ACDC is a public dataset containing short axis view CMR images of 100 patients recorded in NIfTI format [50]. Contour images for the end-systole and end-diastole are provided for each patient [50]. The expert references are manually-drawn on 3D volumes of LV, RV, and myocardial cavities in ED and ES slices [50]. Recordings were performed using two CMR scanners with Siemens Area (1.5T) and Siemens Trio Tim (3T) specifications for 6 years. Cine CMR images were captured during breath-holding with a retrospective or prospective gating and an SSFP sequence in the short axis view [50]. More information on this dataset is provided in [50].

#### 4.1.3. The Kaggle Data Science Bowl Cardiac Challenge

Kaggle is made of 700 datasets for the training and validation phases and 440 datasets for the testing phase [51]. The Kaggle dataset does not provide standard gold LV contours. In addition, the goal and evaluation metric are based on the predicted LV volume at the end-diastole (ED) and end-systole (ES) [51]. More information on this dataset is provided in [51].

#### 4.1.4. Left Ventricle Segmentation Challenge (LVSC)

The LVSC dataset was made publicly available to researchers in MICCAI 2011 [52]. The LVSC dataset contains 200 CMR images of CAD and myocardial infarction patients from several institutions [52]. The primary sequences are cine short-axis steady-state free precession (SSFP) images. Long-axis SSFP cine images are available only for a subset of subjects [52]. The scanners and imaging parameters vary and offer a heterogeneous combination of spatial resolutions of 0.7 to 2.1 mm / pixel and matrix sizes of 156*192 to 512*512. LVSC datasets fall into two groups [52]. The first group includes 100 annotated samples for training and testing. The second group consists of 100 samples without annotation for validation. The gold standard annotations include binary masks delineated by an expert indicating LV myocardium from basal to apical slices for all cardiac phases [52].

#### 4.1.5. Right Ventricle Segmentation Challenge (RVSC)

The RVSC dataset was presented as a section of the 2012 MICCAI workshop [53]. This dataset contains CMR images in DICOM format that have been recorded with the Symphony Tim (1.5T) device. The RVSC dataset consists of manual epicardium and endocardium segments in the ED and ES phases from 48 patients [53]. For this dataset, the data is split as follows: training of 16 patients, test 1 includes 16 cases, and test 2 has 16 patients. For this dataset, ED and ES phases and basal and apical slices have been predefined [53].

#### 4.1.6. CMR Dataset from York University

This dataset contains CMR images in DICOM format with ground truth segmentation of LV endocardial and epicardial [54]. CMR data were recorded from 33 subjects, where each subject's sequence consisted of 20 frames and 8-15 slices along the long axis to make a total of 7980 images [54]. In this dataset, segmentation has been done on images in which both the endocardium and the endocardium of the left ventricle are visible [54]. Thus, there are 5011 segmented CMR images and 10022 contoursin the dataset [54]. Metadata is also available, including pixel spacing, spacing between slices along the long axis, and age and disease of each subject [54].

#### 4.1.7. Left Ventricle Full Quantification Challenge MICCAI 2018 (LVQuan18)

To accurately quantify LV, the STACOM 2018 workshop released the LVQuan18 dataset [55]. The training dataset includes processed SAX MR sequences of 145 subjects [55]. There are 20 frames for each subject. In addition, all ground truth values are provided for each frame [55]. The test dataset contains SAX MR processed sequences of 30 subjects [55]. For each subject, only SAX image sequences of 20 frames without ground truth values are provided [55].

#### 4.1.8. Left Ventricle Full Quantification Challenge MICCAI 2019 (LVQuan19)

The LVQuan19 dataset is the new version of LVQuan18 released at the STACOM 2019 workshop [56]. The training dataset comprises 56 subjects from the processed SAX MR sequences [56]. For each subject, 20 frames are provided and all ground truth values. In the test dataset, the processed SAX MR sequences of 30 subjects are available [56]. At this stage, only the SAX image sequences of 20 frames are provided for each subject, while their ground truth values are not [56].

#### 4.1.9. STACOM

This dataset comprises CMR images of 100 patients with CAD and myocardial infarction [57]. The subjects of this dataset are randomly split into two parts: training and testing [57]. Sixty-six subjects were selected for training, while 34 subjects were used for testing, which resulted in 12,720 training images and 6972 test images [57]. This dataset has a high diverse. Different types of CMR scanners have been employed to record images, and this dataset's CMR image sizes range from 138 x 192 to 512 x 512 pixels [57]. Moreover, each CMR image has a ground truth for the blood cavity and myocardium [57].

#### 4.1.10. STACOM 2017

This dataset was published in a challenge STACOM 2017 for whole heart segmentation. The dataset comprises two parts: training and testing [58]. In the training section, there are 20 CMR images and 20 CT images with ground truth [58]. In the test step, 40 images were provided for each modality without ground truth [58]. CT images are taken from routine cardiac CT angiography and cover the whole heart, extending from the upper abdomen to the aortic arch. The CMR images were acquired using 3D balanced steady-state free precession (b-SSFP) sequences with an acquisition resolution of 2mm in each direction [58].

#### 4.1.11. LASC STACOM 2018

This dataset was part of the STACOM 2018 challenge for LA segmentation [59]. It contains 100 3D LGE-CMRs recorded from patients diagnosed with atrial fibrillation (AF) [59]. A large proportion of data was provided by the University of Utah, while the rest were collected from multiple other institutions [59]. Each 3D D LGE-CMR volume was recorded using a 3.0 Tesla Verio and 1.5 Tesla Avanto scanners. In this dataset, the ground truth binary mask for the LA cavity was annotated by experts for each data [59].

#### 4.1.12. Multi-Modality Whole Heart Segmentation (MMWHS) challenge

This dataset contains 20 CMR data obtained using a Philips Healthcare (1.5T) scanner [60]. The whole heart imaging CMR sequence is balancedsteady-state free precession (b-SSFP) [60]. This database also comprises 20 CT data. CT data were acquired using a Philips Medical Systems scanner [60]. CT images were obtained in axial view, covering the whole heart from the upper abdomen to the aortic arch. More data is provided in the reference [60].

Table 3. Details of dataset used for cardiovascular disease.

| Ref | Dataset | Number of cases | Modality |
|---|---|---|---|
| [49] | SCD | 45 | CMR |
| [50] | ACDC | 100 | CMR |
| [51] | Kaggle | 700 train, 440 test | CMR |
| [52] | LVSC | 200 | CMR |
| [53] | RVSC | 48 | CMR |
| [54] | York University | 33 | CMR |
| [55] | LVQuan18 | 145 train, 30 test | CMR |
| [56] | LVQuan19 | 56 train, 30 test | CMR |
| [57] | STACOM | 66 train, 34 test | CMR |
| [58] | STACOM 2017 | 20 CMR and 20 CT for train, 20 CMR and 20 CT for test | CMR and CT-Scan |
| [59] | LASC STACOM 2018 | 100 3D LGE-CMRs | CMR |
| [61] | UK Biobank | 500,000 | Image, non- image, biological samples, etc. |
| [60] | MMWHS | 20 CMR, 20 CT | CMR and CT-Scan |

### 4.2. Preprocessing techniques

Preprocessing is one of the most substantial steps in CADS for diagnosing heart disease using CMR images. CMR images provide physicians important information about the structure of the heart and assist them in diagnosing CVDs quickly. Though beneficial, CMR data are affected by different artifacts. In addition, CMR images sometimes have low contrast. Hence may lead to inaccurate diagnosis of CVDs from CMR images by specialist physicians. Several preprocessing algorithms have been proposed to address these problems to enhance the performance of DL-based CADS for CVDs detection. Generally, CMR images in DL-based CADS are pre-processed by low-level and high-level procedures.

#### 4.2.1. Low-Level Preprocessing

Low-level techniques are exploited for primary preprocessing of CMR images. The low-level preprocessing plays a significant role in improving CADS performance in CVDs detection. Some of the most important low- level preprocessing techniques comprises of filtering [62], intensity normalization [63], resizing [64], histogram matching [65], cropping [66], segmentation [67], and ROI extraction [68]. Filtering is applied to remove various artifacts from CMR images. Some of the most important filtering algorithms include median and Gaussian filters used in cardiac research [69-70]. Intensity normalization is modifying the range of pixel intensity values and increasing the detection efficiency of CVDs using CMR images [63]. CMR images are normally recorded in high dimensions, so resizing approaches help reduce the CMR dimensions so that they can be fed to the input of DL models [64]. A histogram is another low-level preprocessing technique that aims to enhance the contrast of CMR images [65]. In low-level preprocessing, cropping and segmentation techniques extract important information from CMR images [66]. Then, ROI methods are taken advantage to extract suspected disease areas from CMR images called ROI extraction [68]. Ultimately, the data obtained from ROI extraction is applied to the DL model input. Low-level preprocessing methods are used in all CVDs detection research on CMR using DL techniques.

#### 4.2.2. High-Level Preprocessing

DL models' performance immensely declines when confronted with limited input data. In order to tackle the lack of input data and avoid overfitting, researchers use data augmentation (DA) techniques to increase the training dataset size [71]. Some of the most important DA methods include horizontal flipping and affine transformations rotation and have been investigated in CVDs detection studies [72-73]. GAN models are a new class of DL methods used for DA approaches [74-76]. In [74-76], GAN models have increased input data size.

### 4.3. Deep Learning models

This section describes the most significant DL models used for CVDs detection using CMR images. First, 2D-CNNs [77] and 3D-CNNs [77] are presented with their details. In the following, pre-trained models are introduced, which are a particular mode of CNNs architecture [77]. GAN models [74-76] are an important class of DL models described in this section. U-Net [78-79] and FCN [80] models are two groups of CNNs used for image segmentation applications. A more detailed description of these methods is provided in this section. In another section, AEs and CNN-AEs models [81-82] are introduced to diagnose heart disease. Ultimately, the RNNs and CNN-RNNs models [83-84] are introduced.

#### 4.3.1. Convolutional Neural Network (CNN)

In recent years, DL models have significantly grown in various fields, including medicine [85-90]. CNN architectures have fared exceptionally well in analyzing medical images. Convolutional, pooling, and FC layers are the essential components of a CNN model for the feature extraction and classification [77]. CNN models use supervised learning at the learning stage and encompass different models for classifying and segmenting medical images [91-93]. The most significant advantage of CNN models compared to ML algorithms is feature engineering. The increasing advancement of CNN models has led to increased computational complexity; therefore, hardware resources are also rapidly developing [91-93]. Pre-trained models, 2D-CNNs, and 3D-CNNs are some of the most important CNN models used for classification in the CMR images diagnose CVDs. Further, FCN and U-Net architectures are some of the most popular CNN techniques for CMR image segmentation to diagnose CVDs. Details of CNN models are discussed in the following sections.

**A) 2D-CNN**

In medical approaches, there are often many spatial dependences between the images, making feature extraction difficult [92]. Convolutional layers in CNN models function as spatial filtering [77]. This helps extract useful features by considering spatial dependencies in medical images. Therefore, using convolutional layers lead to automated feature extraction from medical images. The pooling layers in the CNN models function similar to dimension reduction algorithms in ML [77]. Lastly, several fully connected (FC) blocks are used in the last CNN layersto classify input data [77]. , The high efficiency of 2D-CNN models, has led to their massive popularity in studies of identifying CVD from CMR data. Figure (5) illustrates the working of 2D-CNN architecture for CMR image classification to diagnose CVDs.

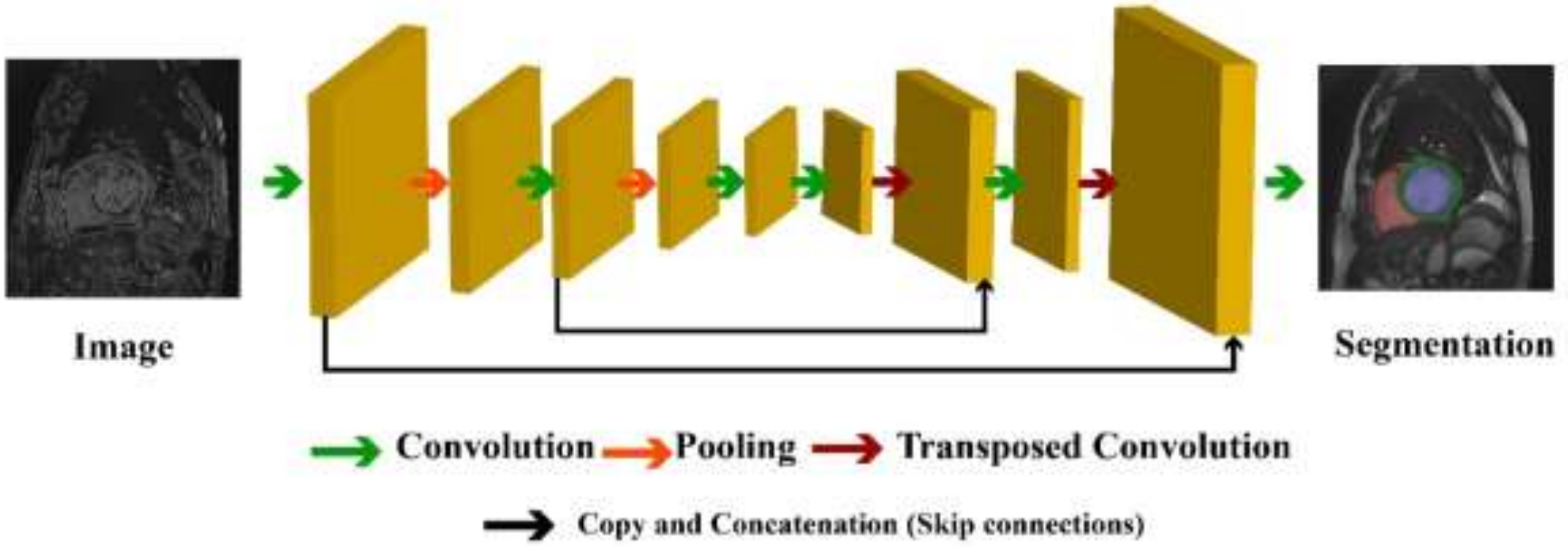

Fig. 5. A typical 2D-CNN for CVDs detection from CMR images.

**B) 3D-CNN**

Medical images such as brain CMR, CMR, CT, and ultrasound are recorded as 3D [94-95]. However, 3D images are highly complex, and it is often challenging for doctors to diagnose diseases based on these data. As a result, researchers have extended 2D-CNN models to 2D-CNNs to obtain more successful outcomes in disease diagnosis from 3D medial images [96-97]. In return, 3D-CNN models require a lot of input data for learning, and researchers often do not have access to datasets with many subjects [96]. Additionally, 3D-CNN models have high computational complexity; hence implementing them requires high-power hardware resources [97]. This always impose challenges in disease diagnosis using 3D-CNN models. Figure (6) illustrates the block diagram of a 3D-CNN architecture for classification of CMR classification for diagnosing CVDs.

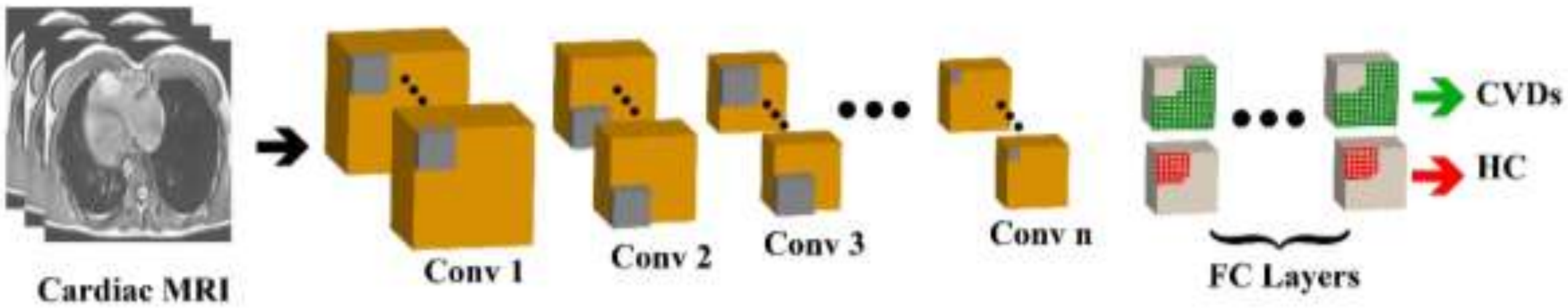

Fig. 6. A typical 3D-CNN for CVDs detection from CMR images

**C) Pretrained Models**

The most significant challenge of studies on disease diagnosis using DL techniques is the lack of access to datasets with many subjects [86]. AI researchers have managed to overcome this challenge in medical studies by proposing deep pre-trained models [98]. Pre-trained models are a group of CNN architectures initially trained on the ImageNet dataset [99-100]. In the following, the weights of the layers have been saved so that researchers can use these architectures to diagnose diseases with fewer subjects [99-100]. For instance, numerous papers have deployed pre-trained models to diagnose CVDs based on CMR images, and researchers have obtained satisfactory results [143]. VGG, AlexNet, etc., are some of the most crucial pre-trained architectures [99-100]. Further, some pre-trained architectures for deep compact size CNNs [101-102] and transformers [103-104] have recently been introduced. Figure (7) illustrates the typical pre-trained model used for CVDs detection from CMR images.

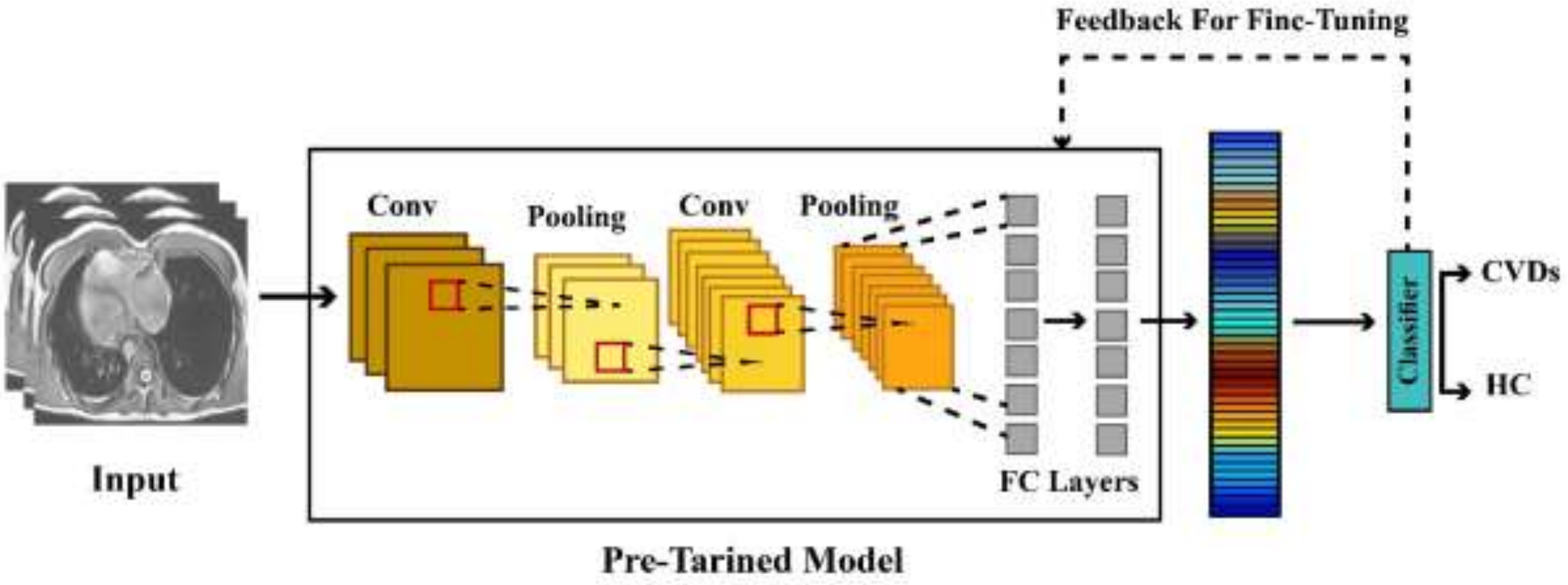

Fig. 7. A typical pre-trained model used for CVDs detection from CMR images.

### 4.3.2. GAN

Generative models have always received much attention due to their ability to model the underlying distribution of data [74-76]. Usually, they consider a simpler family of distributions and try to minimize the KL divergence with the underlying distribution. GANs use a different data generation mechanism, allowing them to create high-quality images. The idea is to create a 2-network minimax game, one network aiming to distinguish between real and fake images and the other trying to fool the first one [74-76]. After training, the generative part (second network) in GAN usually creates realistic data [74-76]. They have been widely used in medical diagnosis tasks, and cardiovascular disease diagnosis. Figure (8) shows the typical GAN model used for CVDs detection using CMR images.

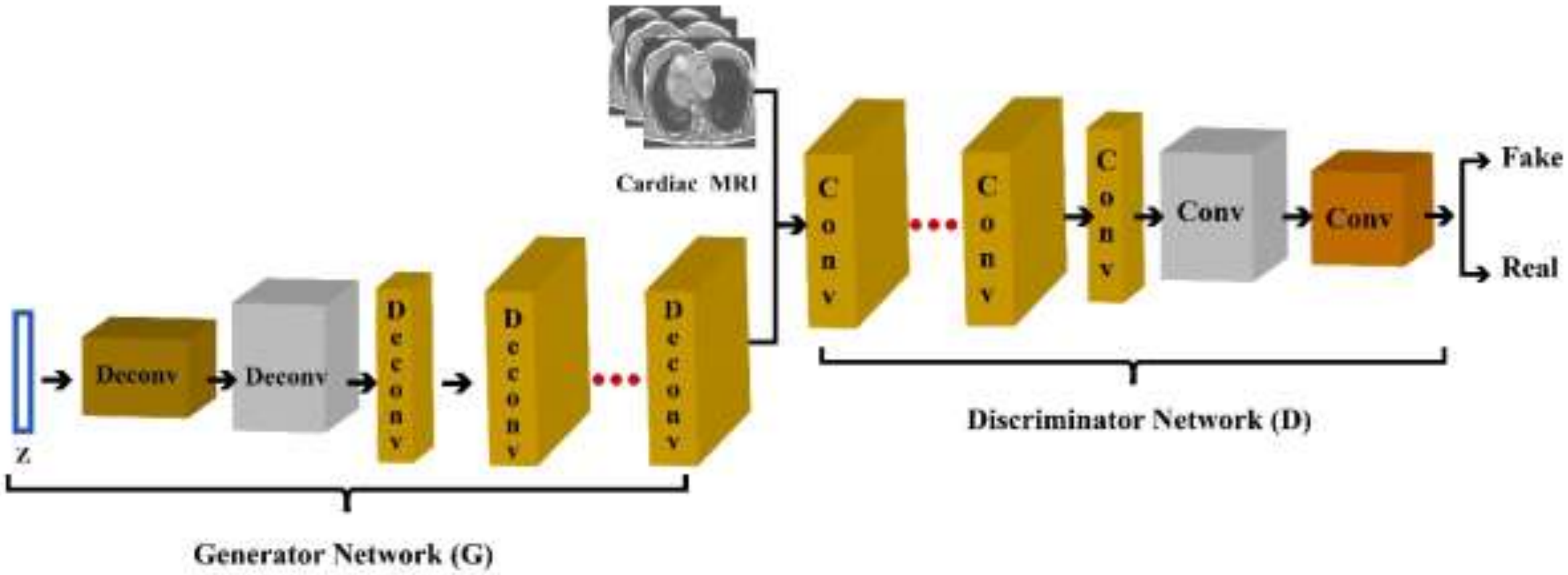

Fig 8. A typical GAN model used for CVDs detection using CMR images.

**4.3.3. FCN**

Long et al. [105] introduced, FCN which is the most fundamental DL architecture used for image segmentation. It is a type of CNN family in which FC layers are not used [105]. Instead, an encoder-decoder structure is used in the FCN architecture for image segmentation [105]. In FCN, the input image is first received with the desired size, then output with the same input dimensions is produced. By applying an image to the input of the FCN model, the encoder first changes the input into high-level feature representation. At the same time, the decoder interprets feature maps and restores spatial details to image space for pixel prediction through a series of convolutional operations and upsampling [105]. Authors in [105] provided more details for the FCN architecture. Figure (9) shows the typical FCN model used for CVDs detection using CMR images..

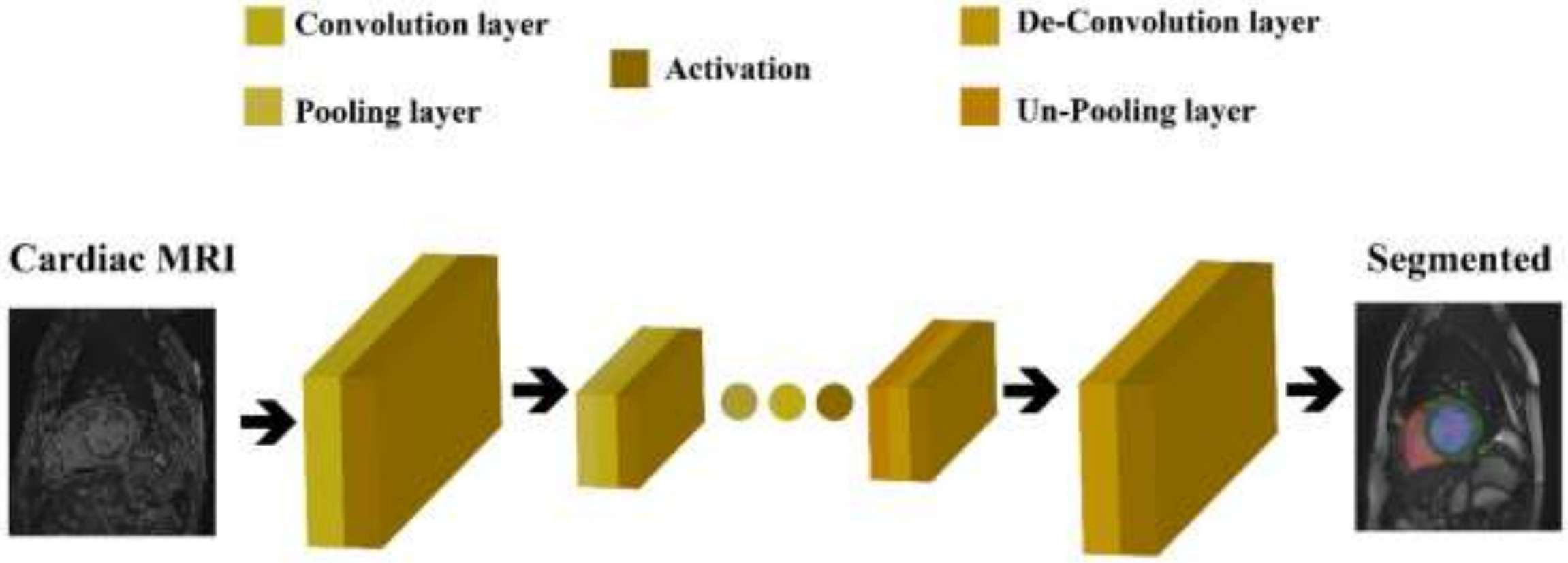

Fig. 9. A typical FCN model used for CVDs detection using CMR images.

**4.3.4. U NET**

Image segmentation is an important step in medical diagnosis, usually used for the localization of different diseases [78-79]. However, conventional CNNs fail to segment; as the image-sized mapping is required for the network's output. FCN and U-net are two famous networks suggested for segmentation [78-79] as they have an encoder-decoder structure that ideally learns required information in the encoder part and encodes it into a latent space, and then decodes that to give a map of the segmented image as the output of the decoder [78-79]. Also, in U-Net, shortcuts between the encoder and decoder are introduced to increase information sharing and help the networks converge faster. These networks have also been used for cardiovascular diagnoses, such as [78-79]. Figure (10) shows the typical U-Net model used for CVDs detection using CMR images.

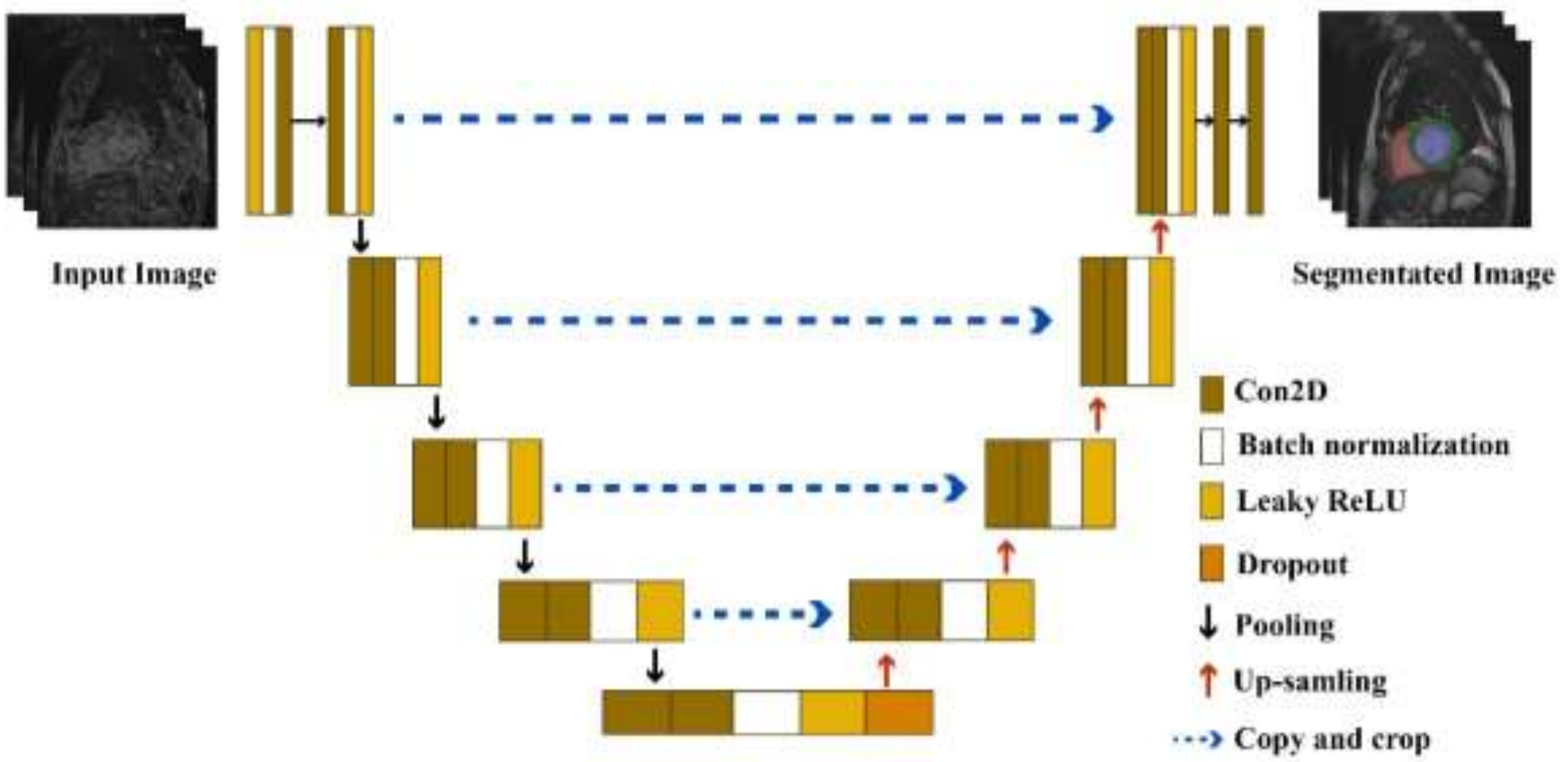

Fig. 10. A typical U-Net model used for CVDs detection using CMR images.

#### 4.3.5. Autoencoder Models

Autoencoders (AE) are the oldest neural networks, but they are still used for many tasks and are even considered state-of-the-art in their domain [77]. The idea is simple; for many tasks, data space is too big, and dimensionality reduction can help dramatically solve the task [77]. Hence, two networks are put back to back, one for encoding the data into a smaller latent space and the other for taking the data back from latent space to the original space, aiming to minimize the re-construction loss [77]. Appropriately trained, AEs should learn to find a robust encoding that preserves the most critical information in data [77] [81-82]. AEs have many different types, such as denoising AEs, Sparse AEs, and Stacked AEs, all aim to resolve one of the challenges AEs face [77] [81-82]. Amongst all types of AEs, Convolutional Autoencoders (CNN-AE) have been used widely in medical diagnosis [106-107]. The idea behind them is to exploit the convolutional abilities by changing the AE layers to convolution and encode spatial patterns into latent space. Figure (11) illustrates the block diagram of a CNN-AE architecture used to diagnose CVDs using CMR images.

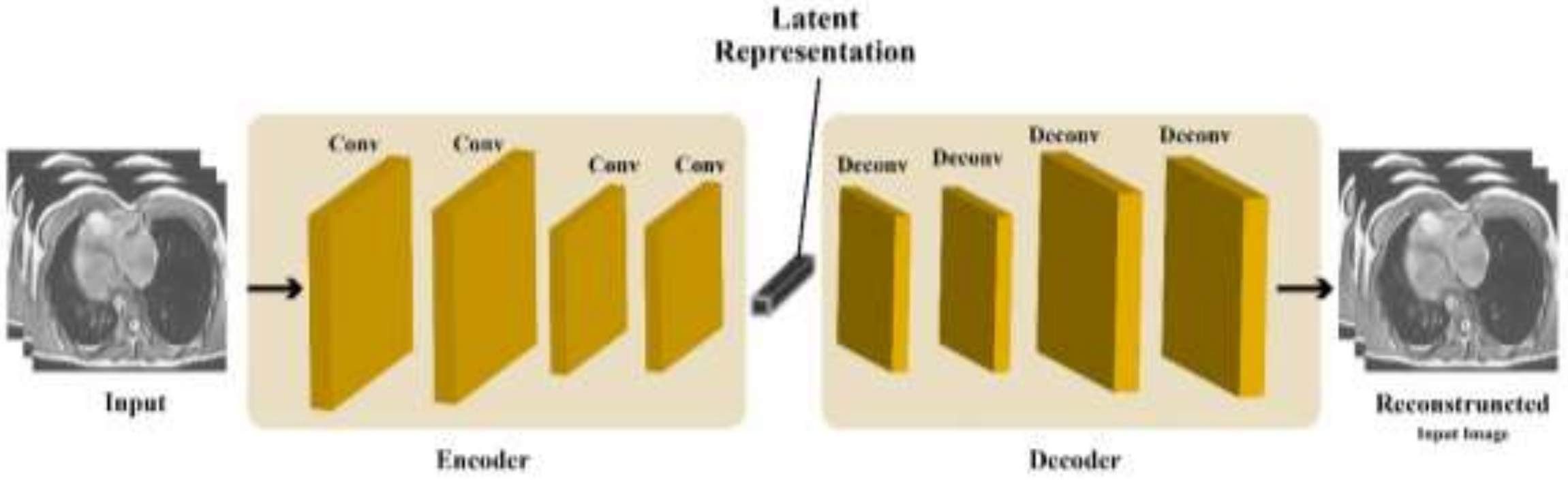

Fig. 11. A typical CNN-AE model for CVDs detection from CMR images

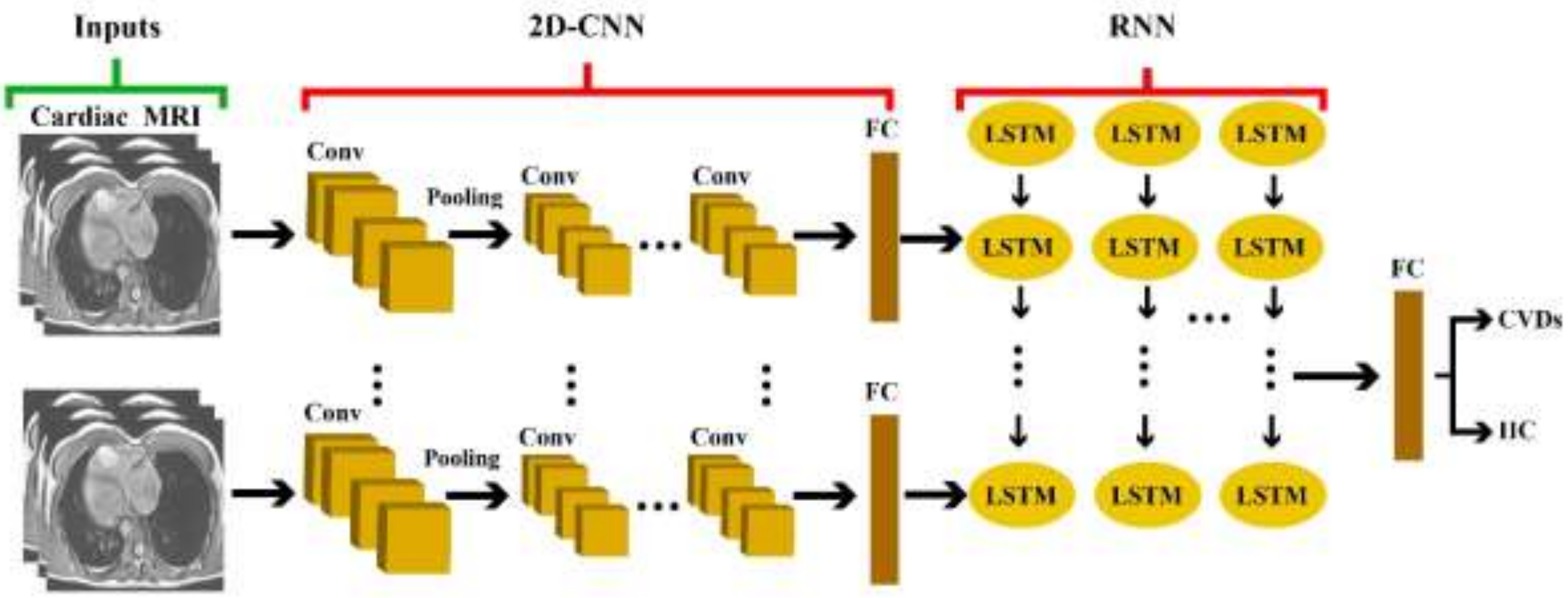

Fig. 12. A typical CNN-RNN model for CVDs detection from CMR images

#### 4.3.6. RNN Models

When applying deep learning models to tasks such as natural language processing, some challenges, such as variable data length or lengthiness, are possible concerns [77]. Amongst these types of challenges, finding temporal patterns is arguably the most important and complicated one, given that these patterns can be of variable length, and previous DL methods have no mechanism to detect them [77]. RNNs, specifically long short-term memory (LSTM) and gated linear unit (GRU) models are built to resolve this issue [77] [83-84], and they are commonly used for signal processing [108], etc. Also, it is common to combine RNNs with other types of networks, such as convolutional ones, to take advantage of both. These models named as CNN-RNN are used to extract spatial and temporal and even Spatio-temporal information from sequential data [77] [83-84]. Figure (12) illustrates a block in the overall diagram of pre-trained CMR data classification architecture to diagnose CVDs.

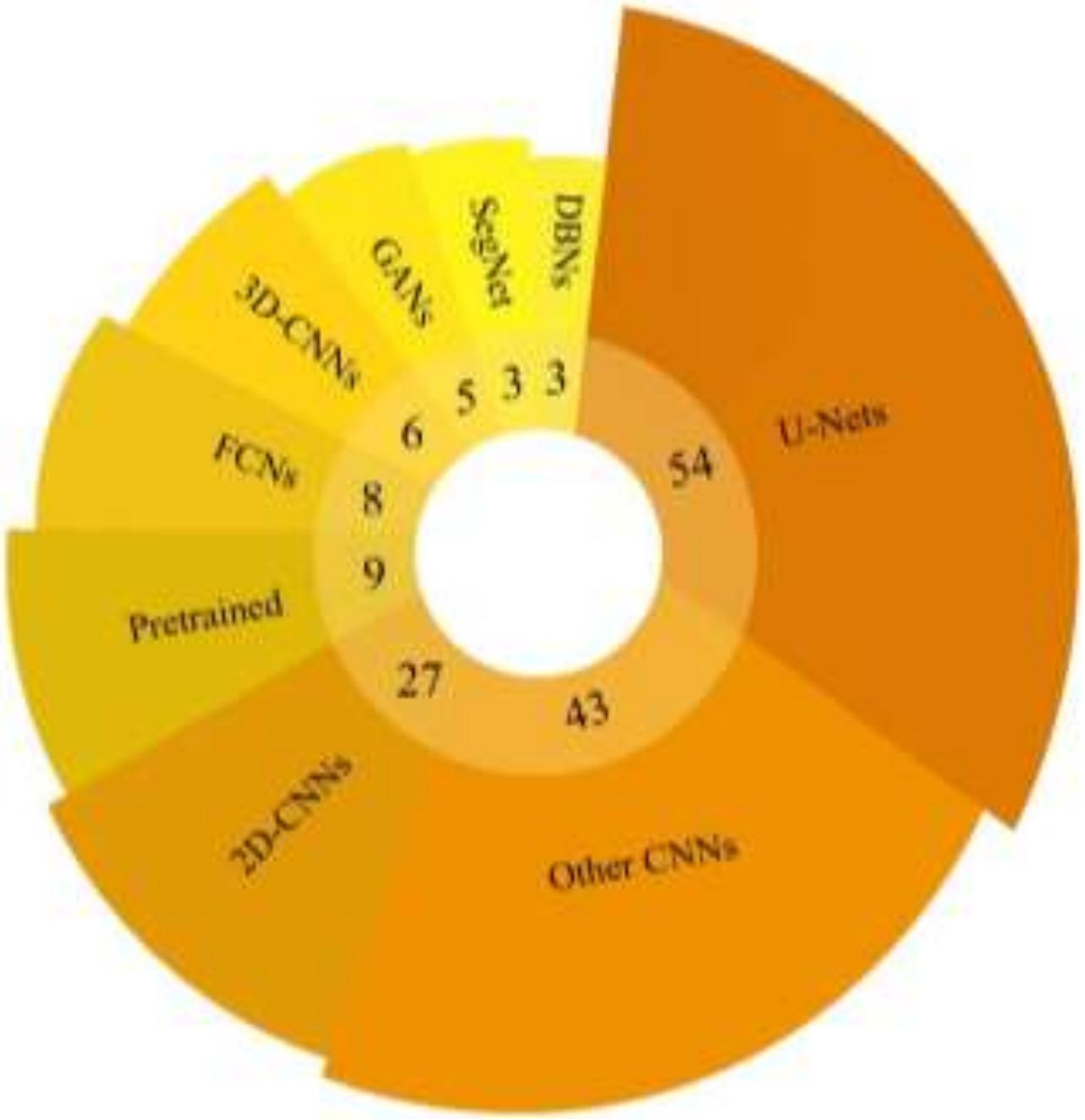

Fig. 13. Various DL models employed in automated segmentation for diagnosis of CVDs in CMR images.

## 4.4. Applications of DL for segmentation of CMR images

As aforementioned, to diagnose CVDs, DL-based segmentation and classification techniques are utilized. The summary of DL-based segmentation works done for CVDs diagnosis are summarized in Table (4). In Figure (13) shows various DL models employed in automated segmentation of diagnosis of CVDs in CMR images. According to Table (4) and Figure (13), CNN models are the most commonly used in CVDs detection using CMR images.

Table 4. Summary of DL-based segmentation works done using CMR image. diagnosis of CVDs using DL methods

| Ref | Application | Dataset | Number of cases | Preprocessing | DNN | Toolbox | Performance |
|---|---|---|---|---|---|---|---|
| [109] | LV | MICCAI 2009 | 45 Subjects | DA | 2D-CNN<br>SAE | NA | DM=94.00% |
| [110] | RV and LV | MICCAI 2012<br>MICCAI 2009 | -- | ROIs Extraction | 2D-CNN<br>SAE | NA | DM=81.00%<br>HD=7.79mm |
| [111] | RV | MICCAI 2012 | 48 Subjects | ROIs Extraction | 2D-CNN | Keras | DM for Endo=86.00%<br>HD=6.9mm<br>DM=84.00%<br>HD=8.9mm<br>EDV R=89.00, ME=7.1<br>ESV: R=84.00, ME=9.6<br>EF R=86.00, ME=7.5 |
| [112] | LV, RV Endo, Epi, and LVM | Sunnybrook<br>LVSC<br>RVSC | 45 Subjects<br>200 Subjects<br>48 Subjects | Different Methods | 2D-CNN | Caffe | Sens=83.00%<br>Spec=96.00% |
| [113] | Multi-slice LV | MICCAI 2009 | 45 CMR | DA | RFCN | NA | Dice=93.50%<br>APD=1.56 |
| [114] | LVM | York University | 33 Subjects | DA | CNN | Caffe | DM=75.00% |
| [115] | Great Vessel | HVSMR 2016 | 20 CMR | DA | Deeply-supervised 3D FractalNet | Caffe | DM=93.00%<br>HD=4.643mm |
| [116] | MYO and blood | MICCAI 2016 | 20 Scans | ROIs Extraction | 2D-Dilated CNN | NA | Blood Pool: DM=93.00%<br>MYO: DM=80.00% |
| [117] | RV | MICCAI 2012 | 48 Subjects | DA | 2D-CNN<br>SAE | NA | DM= 82.50%<br>HD=7.85mm |
| [118] | Myocardial | Sunnybrook | NA | DA, ROIs Extraction | Modified U-Net | Keras | DM=90.00% |
| [119] | LA and PPV | STACOM 2013 | 30 CMR | DA | CardiacNET | TensorFlow | Sens=90.00%<br>Spec=99.00%<br>DM=93.00% |
| [120] | LV | Sunnybrook | 45 Subjects | Random Shuffled | 2D-CNN | Caffe | DM=90.00%<br>HD=5.43mm<br>Sens=90.00%<br>Spec=99.00% |
| [121] | LV, RV and MYO | MICCAI 2017 | 100 Subjects | -- | 2D and 3D-CNN | NA | DM=95.00% |
| [122] | RV and LV Endo and Epi | MHH<br>DSBCC<br>MICCAI 2009<br>RVSC | 502 Subjects<br>1140 Subjects<br>45 Subjects<br>48 Subjects | DA | V-Net | TensorFlow | Different Results |
| [123] | LV | MICCAI 2009 | 45 Subjects | ROI, DA | DBN | NA | ADM=86.00% |
| [124] | LA | 3D LGE-CMR | 60 Subjects | DA | 2D-CNN | TensorFlow | DM=94.20%<br>Sens=91.80% |
| [125] | Shape-Refined Bi-Ventricular | Clinical | Different Subjects | DA | SSLLN | NA | DM:<br>LVC=96.00%, LVW=87.30%<br>RVC=92.90%, RVW=75.50% |
| [126] | Cardiac Bi-Ventricle | Clinical | 145 Subjects | ROIs Extraction | Cardiac-DeepIED (ED+Conv-LSTM) | Keras | Acc LV=99.10%<br>Acc MYO=97.60%<br>Acc RV=98.20% |
| [127] | LV | MICCAI | 45 Subjects | ROIs Extraction | SegNet | MATLAB | -- |
| [128] | LV and RV | UK Biobank | 3078 Subjects | ROIs Extraction | LV-Net | TensorFlow | DM LV-epi=92.30% |
| [129] | MYO and BV | MICCAI 2017 | 150 Subjects | | | | Acc=96.00% |
| [130] | BV | MICCAI 2017 | 150 Subjects | ROIs Extraction | CCGAN | Keras | Different Results |
| [131] | LV | York University | 33 Subjects | ROIs Extraction | 2D-CNN | -- | DM=87.24%<br>Acc=98.39% |
| [132] | Scar | Clinical | 30 Subjects | DA | 2D-CNN | TensorFlow | Acc=96.83%<br>Sens=88.07%<br>DM=71.25% |
| [133] | LV and RV Endo | Clinical | 90 Subjects | Contour delineation, | U-Net | Keras<br>TensorFlow | DM=92.90 %<br>JM=86.90% |
| [134] | LV | MICCAI 2013 | 83 Subjects | Resizing | CapsNet | TensorFlow | DM=94.17 |
| [135] | LV | Clinical | 900 Subjects | Manual Expert Delineations | SegNet | NA | DM Endo=90.00%<br>APD Endo=1.95% |
| | | MICCAI | 45 Subjects | | | | DM Epi=93.00%<br>APD Epi=1.98% |
| [136] | LV | MICCAI 2009 | Different Subjects | NA | DBN | NA | Endo AVP:2.08%<br>Endo ADM:0.90% |
| [137] | LV, MYO, RV | Free-Breathing CMR Data | 12 Subjects | DA, Karhunen-Loeve Transform Filter | U-Net and ResNet | Matlab 2019a | DM LV=91.90%<br>DM MYO=80.60%<br>DM RV=81.80% |
| | | MICCAI 2017 | 150 Subjects | | | | |
| [138] | LV, RV and MYO | MICCAI 2017 | 150 Subjects | Hough Transform, ROIs Extraction, Feature Selection, Feature Scaling, DA | DFCN-C | TensorFlow | DM=91.00%<br>HD=5.43mm |
| | | LV-2011 | 200 Subjects | | | | |
| | | 2015 Kaggle | 500 Subjects | | | | |
| [139] | LV and RV | MICCAI 2017 | 150 Subjects | DA | GridNet-MD | Keras | DM=91.00% |
| [140] | Bi-Ventricle | MICCAI 2017 | 150 Subjects | DA | C-cGANs | Keras<br>TensorFlow | DM LV=96.50%<br>DM RV=94.90%<br>DM MYO=89.30% |
| [141] | LV | Sunnybrook | 45 Subjects | ROIs Extraction, DA | 2D-U-Net | Keras | Different Results |
| | | MICCAI 2017 | 100 Subjects | | | | |
| [142] | LV and RV | Clinical | 100 Subjects | DA | 2D-U-Net | Keras<br>TensorFlow | -- |
| | | MICCAI 2009 | 100 Subjects | | 3D-U-Net | | |
| | | Sunnybrook | 45 Subjects | | DenseNet | | |
| | | MICCAI 2012 | 16 Subjects | | | | |
| [143] | LV and RV | MICCAI 2017 | 100 Subjects | DA | Proposed Method | PyTorch | DM RVC=90.30%<br>DM LVM= 89.20%<br>DM LVC=94.20% |
| | | Clinical | 100 Subjects | | | | HD RVC=13.830mm<br>HD LVM= 8.786mm<br>HD LVC=6.641mm |
| [144] | LV | Clinical | 100 Subjects | -- | FC- U-net | PyTorch | DM Epi=96.00%<br>DM Endo=94.00% |
| [145] | LV | York University | 30 Subjects | -- | U-Net and GoogleNet | Keras | DM=89.00% |
| [146] | LV | MICCAI 2011 | Different Subjects | DA | 2D-CNN | TensorFlow | DM=88.00% |
| | | MICCAI 2009 | 100 Subjects | | | | |
| [147] | LV | Sunnybrook | 45 Subjects | -- | FR-net | Caffe | DM=93.00%<br>APD=1.41 |
| [148] | LV | MICCAI 2017 | 150 Subjects | ROIs Extraction, DA | 2D-CNN2 | PyTorch | DM LV ED= 96.00%<br>DM LV ES= 92.00% |
| | | | | | | | DM MYO ED=88.00%<br>DM MYO ES=89.00% |
| [149] | Cardiac Walls | Clinical | 20 Subjects | DA | PC-U Net | -- | DM=88.50%<br>HD=7.050mm |
| [150] | LV | Clinical | 33 Subjects | CLAHE | DT-GAN | PyTorch | HD=2.23mm<br>DM=93.00% |
| [151] | CMR | MICCAI 2017 | 100 Subjects | DA | DBAN | NA | DM=ED 96.00%<br>DM ES=90.00% |
| | | | | | | | HD ED=6.7mm<br>HD ES=8.1mm |
| [152] | Scar | Clinical | 155 Subjects | DA | ACSNet | Keras<br>TensorFlow | Acc LV=96.00% |
| | | MICCAI 2017 | 245 Subjects | | | | |
| | | Sunnybrook | | | | | |
| [153] | MYO | MICCAI 2020 | 150 Subjects | -- | 3D U-Net | PyTorch | DM MYO=87.86% |
| [154] | LV, RV, and MYO | MICCAI 2017 | 150 Subjects | ROIs Extraction, YOLOv3 | LFCN | TensorFlow | DM LV ED=96.00%<br>DM LV ES=91.00%<br>DM RV ED=93.00%,<br>DM RV ES=85.00%<br>DM MYO ED=87.00%<br>DM MYO ES=89.00% |
| [155] | Cardiac Multi-task | MyoPS 2020 | 45 Subjects | DA | CMS-U-Net | PyTorch | DM MYO=58.10% |
| [156] | LV, RV, and MYO | Clinical | 175 Subjects | DA | 2D-CNN | NA | Acc=97.60% |
| [157] | LV blood pool and MYO | MICCAI 2020 | 150 Subjects | -- | U-Net | NA | Acc=92.00%<br>DM= 86.28% |
| [158] | LVM, LV, and RV | Different datasets | 350 Patients | DA | U-Net | NA | DM= 85.48% |
| [159] | | 36 Unique Datasets | 32 Subjects | DA | U-Net | Keras TensorFlow | DSC=88.00% |
| [160] | LV and RV | Clinical | Different Subjects | -- | U-Net | NA | DM=95.00 |
| [161] | LA, LV, RV Endo, and MYO, at ED and ES | STACOM | 100 Subjects | GCAM | DR-U-Net | Keras TensorFlow | DM =92.80%<br>HD=20.3mm<br>ASD=1.38mm |
| [162] | LV,RV, and MYO | MICCAI 2017<br>UK Biobank | 100 Subjects<br>100 Subjects | DA | U-Net | NA | DM=93.50% |
| [163] | RVM and LVM | MICCAI 2017 | 1902 Cardiac MR Images | DA | 2D-CNN | Keras TensorFlow | DM=91.60% |
| [164] | LV Cavity, MYO, and RV Cavity | UK Biobank | 100 Subjects | DA | 2D-CNN | Python, Theano | DM LV=92.00%<br>DM MYO=85.00%<br>DM RV=89.00% |
| [165] | LV, RV, and MYO | MICCAI 2017 | 150 Subjects | DA | 2D-CNN<br>3D-CNN | PyTorch | Different Results |
| [166] | LV | -<br>Kaggle | 45 Subjects<br>500 Patients | DA | 2D-CNN | Theano | CRPS =0.084<br>RMSE =65.6 |
| [167] | AS | Clinical | 20 Subjects | Different Methods | SSAE | -- | AUC = 94.00% |
| [168] | LVM | Clinical | 8 Subjects | -- | ResNet-56 | MXNet | DM= 86.00%<br>HD=4.01mm |
| [169] | LV | LVSC<br>Kaggle | 200 Subjects<br>1140 Subjects | DA | CPL Network<br>MB Network | Python, TensorFlow | Sen=88.00%<br>Spec=95.00% |
| [170] | LV Cavity | York University<br>MICCAI 2009 | 33 Subjects<br>45 Subjects | -- | U-Net | Keras, Theano | DM=93.00% |
| [171] | VS | Automated Cardiac Diagnosis Challenge 2017 | 100 Subjects | DA | 3D FCN | TensorFlow | DM=82.27%<br>Prec=89.81% |
| [172] | LA, PV, and AFS | Clinical | ??? | Different Methods | SSAE | NA | Acc=91.00%<br>Sen=95.00% |
| [173] | LV, RV And LV | MICCAI 2017 | 100 Exams | -- | 3D-CNN | NA | DM=90.00%<br>HD=10.4mm |
| [174] | LV | Second Annual Data Science Bowl | 7 Subjects | -- | 2D-CNN | MXNet | HD=3.70mm |
| [175] | LV, LVi, and RV | Clinical | -- | DA | FastVentricle | Keras, TensorFlow | Different Results |
| [176] | LV | SCD<br>MICCAI 2017<br>Kaggle | -- | DA | U-Net | NA | -- |
| [177] | | Clinical | 1,912 Subjects | DA | CVAE | TensorFlow | DM=87.92% |
| [178] | RV | Clinical | 26 Subjects | -- | 3D CNN | -- | DM=82.81% |
| [179] | RV | MICCAI 12 | -- | -- | Multi-Task DNN | TensorFlow | DM=87.20% |
| [180] | LV | Clinical | 30 Subjects | -- | U-Net | PyTorch | DM=94.00% |
| [181] | MYO | Clinical | 348 Subjects | -- | RSE-Net Model | PyTorch | DM=82.01% |
| [182] | LA and PV | Clinical | ??? | -- | Different Models | TensorFlow | Acc=99.70%<br>DM=89.70% |
| [183] | | UK Biobank | 220 Subjects | DA | ??? | NA | Different Results |
| [184] | Scar | Clinical | Subjects | DA | Modified Version of ENet | NA | Acc=97.00%<br>Sen=88.00% |
| | | | | | | | DSC=71.00% |
| [185] | LV | TWINS-UK | 68 Subjects | -- | T-FCNN | NA | DM=98.15% |
| [186] | LV, MYO, and RV | UK Biobank | 5000 Images | DA | Syn-net<br>LI-net | PyTorch | Different Results |
| [187] | Atrial | Clinical | 3 Subjects | DA | Modified U-Net | NA | DM=94.88%<br>HD=7.56mm |
| [188] | LV and MYO | Clinical<br>MICCAI 2018<br>MICCAI 2019 | 75 Subjects<br>145 Subjects<br>56 Subjects | DA | 2D-CNN | NA | -- |
| [189] | LV | MICCAI 2009 | 45 Subjects | -- | 2D-CNN<br>U-Net | NA | DM=95.10%<br>HD=3.641mm |
| [190] | LV | SCD | 45 Images | -- | 2D-CNN | NA | Acc=94.00%<br>Sen=94.11%<br>DM= 94.00% |
| [191] | BV | Clinical | 145 Subjects | -- | Bi-DBN | Theano | different Results |
| [192] | LV | STACOM<br>MICCAI 2017 | 100 Subjects<br>100 Subjects | -- | OF-net | NA | APD=0.90<br>DM =95.00% |
| [193] | LV | MICCAI 2019 | 56 Subjects | DA | CNN | Keras, TensorFlow | DM Epi=96.10%<br>DM Endo=94.90%<br>DM MYO=86.70% |
| [194] | LV, LV, RV | MICCAI 2017<br>MICCAI 2017 | 150 Subjects | -- | 2D-CNN | NA | DM=90.00% |
| [195] | LV | MICCAI 2017 | 150 Subjects | -- | SegAN + U-Net B | NA | DM=95.87% |
| [196] | LV,RV, and MYO | Clinical | 45 Subjects | DA | SRSCN | TensorFlow | DM MYO=81.20%<br>DM LV=91.50%<br>DM RV=88.20% |
| [197] | Multi-Sequence | MICCAI 2019 | 45 Subjects | DA | Dilated Residual U-Shape Network and CNN | Keras | DM LV=82.40%<br>DM MYO=61.00%<br>DM RV=71.00% |
| [198] | Segmentation | MICCAI 2017<br>BraTS 2017 | 150 Subjects<br>289 Subjects | DA | Weighted-RNN-GAN | Keras, TensorFlow | Different Results |
| [199] | LA | Clinical | 100 Subjects | Resizing | CNN Inception V4 with AE | Keras, TensorFlow | DM=93.10%<br>HD=4.2mm |
| [200] | LA and RV | Different Dataset | Different Subjects | ROIs Extraction | 2D CNN, SAE | NA | Acc=98.66% |
| [201] | LV,RV, and MYO | MICCAI 2017 | 150 Subjects | ROIs Extraction | 2D Residual CNN<br>Bi-CLSTM | NA<br>NA | Different Results |
| [202] | MYO | Clinical | 195 Subjects | -- | CNN | NA | DM=81.37% |
| [203] | LV, RV, and LVM | MICCAI 2017 | 100 Subjects | ROIs Extraction | L-CO-Net | NA | DM LV=96.80%<br>DM RV=93.30%<br>DM LVM=89.50% |
| [204] | Left and Right Chamber | Clinical | 210 Subjects | -- | U-Net | NA | Different Results |
| [205] | BV | MICCAI 2017 | 100 patients | DA | U-Net | PyTorch | DM RVC= 79.60%<br>DM LVM=84.60%<br>DM LVC=90.80% |
| [206] | LV | Sunnybrook | 800 image slices | -- | CNN with U-Net | NA | F1-S=95.90% |
| [207] | RA | Clinical | 550 Images | DA | U-Net | NA | DM=94.88%<br>JM=90.33%<br>HD=7.5625mm |
| [208] | RA | Clinical | 242 Subjects | -- | U-Net | NA | HD=4.64mm |
| [209] | LV | InCor<br>Sunnybrook<br>MICCAI 2017<br>MICCAI 2019<br>LVSC 2011 | 59 Subjects<br>45 Subjects<br>150 Subjects<br>56 Subjects<br>200 Subjects | -- | U-Net | NA | DM Endo=82.00%<br>DM Epi=86.00%<br>HD Endo=5.81mm<br>HD Epi=6.69mm |
| [210] | LV and RV | Clinical<br>MICCAI 2017 | 63 Subjects<br>100 Subjects | DA | U-Net | Keras, TensorFlow | Acc=97.00% |
| [211] | LV Blood-Pool, MYO and RV Blood-Pool | MICCAI 2017 | 100 Subjects | DA | SegAN and 2D U-Net | NA | Different Results |
| [212] | MYO Infarction | Infarction Segmentation Challenge | 15 Subjects | Different Methods | Chained U-Net | NA | DM=32.00% |
| [213] | MYO | Clinical | 355 Subjects | -- | U-Net | TensorFlow Keras | DM MYO=94.34% Acc MYO=99.873% |
| [214] | LV | Clinical | 42 Female Breast Cancer Datasets | SFP, Zero-Crossing Edge Detection | DeepLabV3+ DCNN | NA | Acc=97.00% Dice=89.00% |
| [215] | MYO | HVSMR 2016 | 20 MR Images | -- | 3D FCNN | Keras | DM MYO=76.20% |
| [216] | LV, RV, and MYO | Clinical | 150 Subjects | ROIs Extraction, DA | U-Net | TensorFlow | DM=92.00% HD=12.18mm Acc=92.00% |
| [217] | LV | Clinical | 150 Subjects | ROIs Extraction, DA | 2D FCN | MATLAB | DM= 93.00% Sen=98.00% Spec=94.00% |
| [218] | LV,RV | UK Biobank | 1000 Subjects | -- | CNN | NA | -- |
| [219] | LV, RV, and MYO | MICCAI 2017 | 150 Patients | U-Net For ROIs Extraction | FCN, U-Net | Keras TensorFlow | DM LV=96.30% |
| [220] | LV, RV | MICCAI 2017 | 100 Subjects | -- | U-Net | NA | Acc=90.00% |
| [221] | LV | MICCAI 2009 | 45 Subjects | -- | LsU-Net | TensorFlow | DM LV Endo=92.15% DM LV Epi=95.42% |
| | | 2012 RV Segmentation Challenges | 16 Subjects | | | | |
| | | Clinical | 17 Subjects | | | | |
| [222] | LV Myocardium | Clinical | 56 Subjects | Contrast Enhancement | 2D-Residual Neural Network | PyTorch | DM=85.43% |
| [223] | Four Cardiac Chambers | Clinical | 150 Subjects | DA | U-Net CNN | MATLAB | DM=89.00% |
| [224] | LV, RV-LV | Clinical | 108 Subjects | DA | U-Nets | TensorFlow | DM=87.00% HD=5.9mm |
| [225] | Blood Pool And Myocardium | HVSMR | 10 Subjects | NA | GCEFG-$R^2$Net | PyTorch | DM Blood Pool=95.80% DM MYO=83.60% |
| [226] | LV and MYO | Kaggle | 1140 Subjects | DA | U-Net (ResNet34 Backbone) | PyTorch, FastAI | DM LV=90.00% DM MYO=79.10% |
| | | Clinical | 22 Subjects | | | | |
| [227] | LV, RV, and MYO | imATFIB | 20 Cases | ROIs Extraction, DA | U-Net, DeepLabV3+ | NA | DM=92.90% |
| [228] | LV | Sunnybrook | 805 Images of 45 Cine-CMR | -- | ROR-Unet | Keras | Different Results |
| [229] | FCEA Tissue | Clinical | 100 Subjects | DA | U-Net | TensorFlow | DM=77.00% |
| [230] | Pathology | MICCAI 2020 | 45 Subjects | DA | TAU-Net | PyTorch | DM=63.60% |
| [231] | RVC,LVC,LVM | MICCAI 2017 | 150 Subjects | ROIs Extraction | U-Net | TensorFlow | Acc=92.00% Acc=91.00% |
| [232] | LV | Clinical | NA | ROIs Extraction | U-Net | NA | DM Epi= 94.07% DM MYO=88.27% DM Endo=91.77% |
| [233] | LV, RV | DBI | Different Subjects | ROIs Extraction, DA | U-Net | TensorFlow | DM LV=96.1% |
| [234] | LV | MICCAI 2017 | 100 Subjects | Data Enhancement | MMNet | PyTorch | DM=95.10% HD=7.00mm |
| [235] | RV | Clinical | 45 Subjects | DA | FCDL | TensorFlow | DM=87.00% HD=7.55mm |
| [236] | Different Segmentation | 2019 MSCMRSeg | 45 Subjects | DA | Proposed Method | NA | Different Results |
| [237] | LV,RV, and MYO | MICCAI 2017 | 150 Subjects | DA | Proposed Method | PyTorch | ED: DM LV=89.67% DM RV=81.46% DM MYO=72.60% |
| | | | | | | | ES DM LV=81.33% DM RV=70.80% DM MYO=76.56% |
| [238] | | MICCAI | | -- | ACSNet | | DM LVM=79.00% |
| | MYO And Scar/Fibrosis | Cardiac MR LV Segmentation Challenge | Different Subjects | | | Keras TensorFlow | HD LVM=6.70mm |
| [239] | RV | RVSC<br>MMWHS | 48 Subjects<br>80 Subjects | DA | TSU-net | Keras TensorFlow | DM Endo=86.00%<br>DM Epi=90.00% |
| [240] | CMR | MM-WHS Challenge 2017 | 60 CMR,60 CT Images | GAN | GBCUDA | NA | DM=59.20% |
| [241] | LV | Clinical | 33 Subjects | ROIs Extraction | GAN, U-Net | PyTorch | DM=96.97% |
| [242] | LV | LVSC | 200 Subjects | -- | Attention U-Net | Keras TensorFlow | Sen=87.00%<br>Spec=92.00% |
| [243] | CMR | Clinical | 3333 Frames | -- | MIFNet | PyTorch | DM=97.23%<br>Sen=93.55% |
| [244] | LV | MICCAI 2017 | 150 Subjects | DA | FCN-MSPN and Co-Discriminators | Keras TensorFlow | Different Results |
| [245] | MYO | MyoPS 2020 | 45 Subjects | DA | AWSnet | PyTorch | DM=72.00% |
| [246] | LV, RV, and MYO | York University<br>MICCAI 2009<br>MICCAI 2017 | Different Subjects | Gamma Transformation | BLU-Net | PyTorch | Different Results |
| [247] | CMR Images | MICCAI 2018 | 145 Subjects | ROIs Extraction | U-Net, MC-Seg | PyTorch | DM=88.60%<br>HD=4.21mm |
| [248] | CMR Images | MICCAI 2009<br>MICCAI 2017 | 45 Subjects<br>100 Subjects | -- | U-Net Backbone | Keras, TensorFlow | DM LV=93.41%<br>DM RV=89.74%<br>DM MYO=89.74% |
| [249] | LV, RV, and MYO | MICCAI 2017 | 100 Subjects | DA | U-Net | NA | DM LV=72.00%<br>DM RV=53.00%<br>DM MYO=69.00% |
| [250] | LV, Scar | Clinical | 34 Subjects | DA | CTAEM-Net | Keras a TensorFlow | Acc LV=86.43%<br>Acc Scar=90.18% |
| [251] | LA | LASC | 20 Subjects | -- | 3D SR-Net | NA | DM=93.29%<br>F1-S=82.37% |
| [252] | LV | MICCAI 2009 | 45 Subjects | DA | FCDA-Net | PyTorch | DM ED= 93.59%<br>DM EP=94.81%<br>HD ED= 4.95%<br>HD EP=3.18% |
| [253] | MYO | Clinical | 60 Subjects | -- | U-Net, Dense Nets, Attention Nets | Keras, TensorFlow | Different Results |

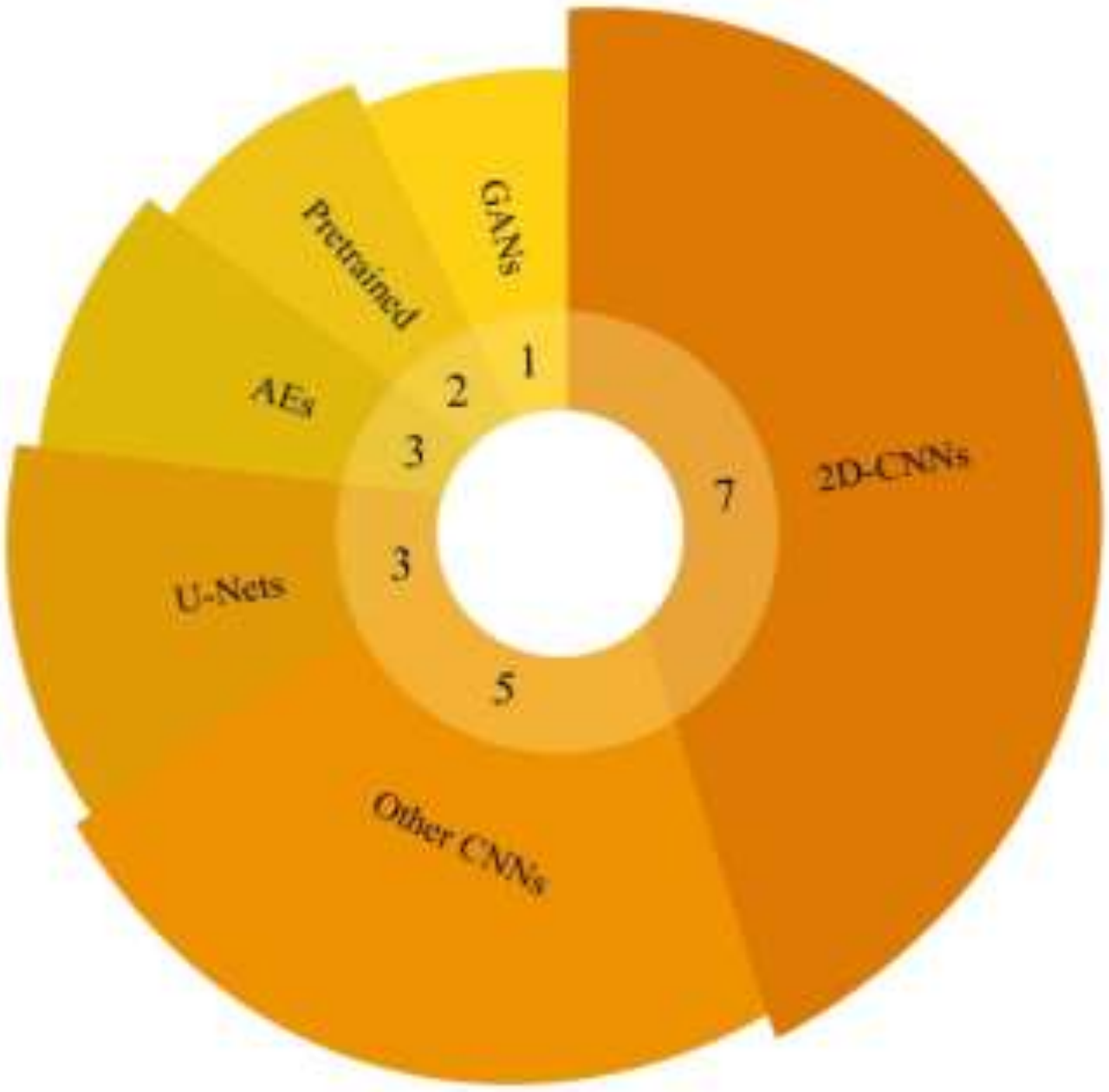

Fig. 14. Deep learning methods in classification of CMR images for Diagnosis of CVDs

## 4.5. Applications of DL for classification of CMR images

In this section, CVDs diagnosis papers using DL classification models are presented. A primary objective of these papers is to diagnose CVDs from HC. A summary of papers about CMR-based CVD diagnosis using DL classification models is presented in Table (5). Additionally, the number of DL classification models used to diagnose CVDs is depicted in Figure (14). As shown in Figure (14) and Table (5), CNN models are most exploited in classifying CMR images to diagnose CVDs. CNN models have performed extremely well in various medical applications to diagnose CVDs.

Table 5. Research in classification of CMR imagers for in diagnosis of CVDs using DL methods

| Ref | Application | Dataset | Number of cases | Preprocessing | DNN | Classifier | Toolbox | Performance |
|---|---|---|---|---|---|---|---|---|
| [254] | LV | ADSB | 937 Subjects | Gabor Filter | 2D-CNN | FC | Keras | Different Results |
| [255] | End-Diastole and End-Systole Frames | Clinical | 420 Subjects | DA | TempReg-Net | FC layer | Caffe | ADF ED=0.38<br>ADF ES=0.44 |
| [256] | LV | DSBCCD | 1140 Subjects | ROIs Extraction, DA | 2D-CNN | FC Layer | Keras | -- |
| [257] | Classification | ILSVRC 2012 | 215 Subjects | DA | CaffeNet, CardioViewNet | Different | Caffe | F1-S=97.66%<br>Recall=97.62% |
| [258] | MYO Ischemia | MICCAI 2009 | 21 Subjects | ROIs Extraction | CNN | NA | NA | Acc=86.39%<br>Sen=90.00% |
| [259] | End-Diastole and End-Systole Frames | Free-Breathing CMR Data<br>STACOM2011 | 10 Subjects<br>200 Subjects | DA | 2D-CNN | NA | Caffe | Acc=76.50% |
| [260] | Heart and right ventricle | Clinical<br>York University | 65 Subjects<br>33 Subjects | DA | NF-RCNN | Softmax | -- | AUC=98.00%<br>Recall=96.00% |
| [261] | LV | DSBCCD | 1140 Subjects | ROI Extraction | CNN | FC layer | Keras | EDV $R^2$=97.40<br>ESV $R^2$=97.60<br>EF $R^2$=82.80 |
| [262] | LV | Sunnybrook, Kaggle | 1140 Subjects | LBP Cascade Detector, DA | HFCN | Softmax | -- | RMSE=13.20<br>ESV RMSE=9.31 |
| [263] | Dense Thickness Estimation | MICCAI 2017<br>Synthetic<br>MICCAI 2019<br>MS-CMRSeg | 100 Subjects<br>Different Subjects | -- | U-Net-k | NA | NA | MSE=14.30<br>MAE=28.50 |
| [264] | Detection | Clinical | 363 Subjects | BBoxes, Visualization | 2D-CNN | NA | NA | AUC=89.10% |
| [265] | Detection | Clinical | 350 Subjects | ROIs Extraction | 2D-CNN | Softmax | Keras | Acc=94.84%<br>Sen=92.73%<br>Spec=94.27% |
| [266] | Classification | MICCAI 2017 | 150 Subjects | Feature Extraction | Modified 2D and 3D U-Nets | Ensemble Learning | -- | Acc=92.00% |
| [267] | Classification of MYO | Clinical | 200 Subjects | -- | Pretrained Models | Softmax | -- | Acc=82.10% |
| [268] | Classification and Prediction | Clinical | 198 HCM Subjects | -- | DeeplabV3<br>InceptionResnet V2 | LSTM Model | Keras | Different Results |
| [269] | Cardiac view | Clinical | Different Subjects | -- | AE | Softmax | -- | Acc=96.70% |
| [270] | Multitype cardiac indices estimation | Clinical | 145 Subjects | ROIs Extraction | DCAE | ????? | Caffe | NA |
| [271] | LV | Clinical | 26 Subjects | ROIs Extraction | AE | -- | -- | Acc=97.50%<br>Sens=84.20%<br>Spec=98.60% |
| [272] | MYO | Clinical | 566 Subjects | -- | CNNEC | FC layer | -- | Acc=95.30% |
| [273] | Cardiac Contraction | UK Biobank | 12000 Subjects | -- | CGAN | Softmax | -- | DM=89.00% |

## 4.6. Applications of DL for CMR images (other approaches)

This section discusses other applications of DL on CMR images. Table (6) presents DL applications using CMR data. Some of the most important applications of DL on CMR images comprises reconstruction [294], automatically computing cardiac views, generating CMR slices, and motion artifact correction. The number of DL models used for this are displayed in Figure (15). As can be seen, CNN models have gained great popularity in medical imaging.

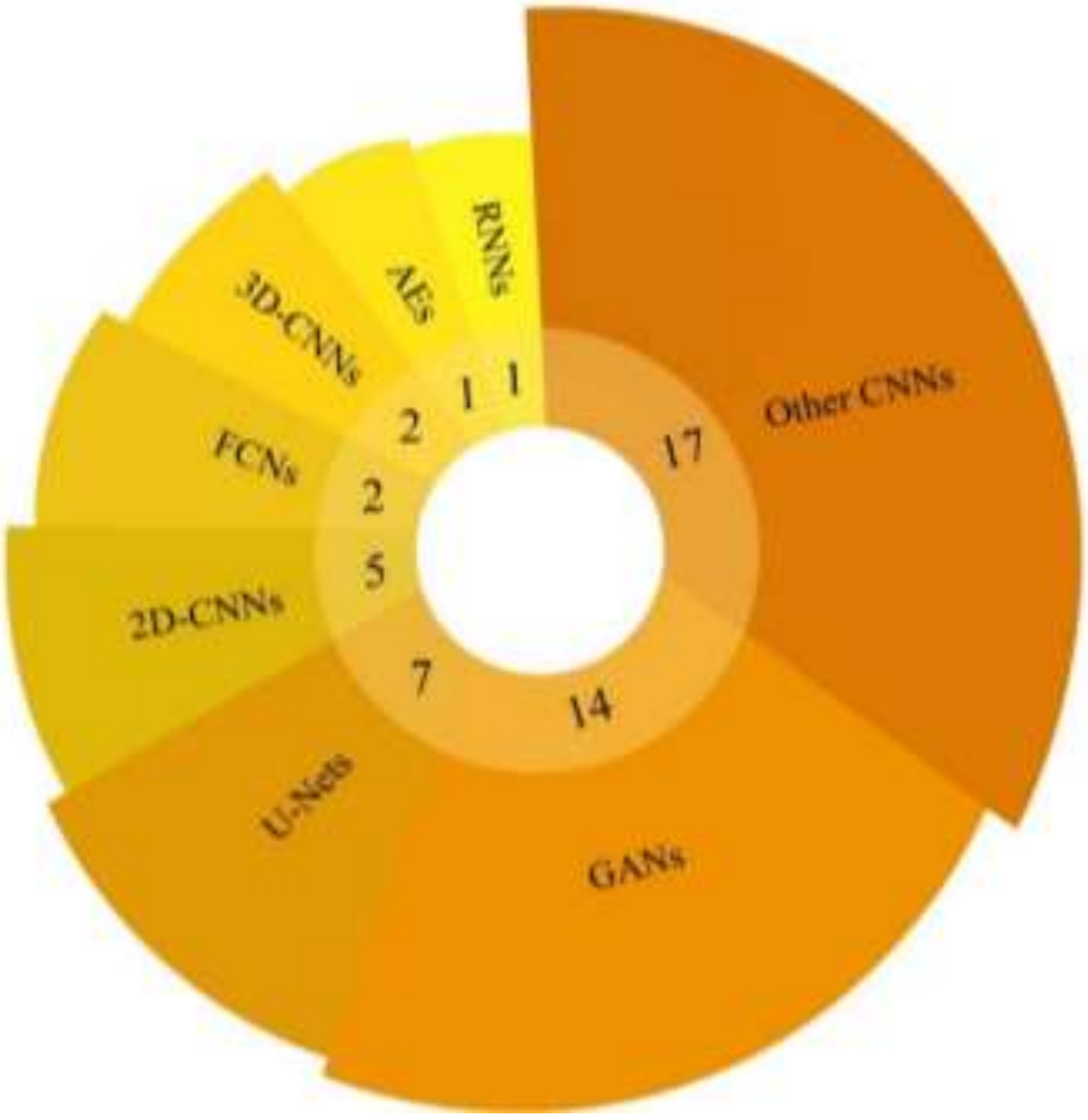

Fig. 15. Deep learning methods used for different approaches in CMR images.

Table 6. Deep leering models used in CMR images for different approaches.

| Ref | Application | Dataset | Number of cases | Preprocessing | DNN | Classifier | Toolbox | Performance |
|---|---|---|---|---|---|---|---|---|
| [274] | Identifying the Missing Apical and Basal Slices | UK Biobank | 100 Subjects | Global mask | 2D-CNN | Softmax | NA | Pec=81.61% Pec=88.73% |
| [275] | Incomplete LV Coverage | UK Biobank | 3400 Subjects | -- | SCGANs | SVM | TensorFlow | Ac=92.50% Prec=87.60% Rec=90.50% |
| [276] | MRCA | Clinical | 10 Subjects | ROIs Extraction | DLR | -- | NA | DLR-HR-MRCA:11.3 |
| [277] | Left Ventricle | York University | 33 Subjects | Pyramid of Scales | 2D-CNN | Softmax | NA | Acc=98.66% Sen=83.91% Spec=99.07% |
| [278] | LV and RV on short-axis CMR Images, LA, and RA | UK Biobank | 4875 Subjects | DA | 2D-CNN | Softmax | TensorFlow | Different Results |
| [279] | LV classification | SCD<br>CAP | 140 Subjects | DA | 2D-CNN | Softmax | Keras | DM=90.00% |
| [280] | Identifying | Clinical | Different Subjects | DA, ROIs Extraction | U-Net | -- | NA | DM=76.00% |
| [281] | LV | STACOM18 LVQuan | 145 Subjects | DA | U-Net | -- | PyTorch | Different Results |
| [282] | Cardiac Cavity Segmentation Task | MICCAI 2017 | 200 CMR Images | DA | XCAT-GAN | -- | NA | Different Results |
| | | York University | 33 CMR Images | | | | | |
| | | XCAT Simulated | 66 CMR Images | | | | | |
| | | SCD | 45 CMR Images | | | | | |
| | | Clinical CMR | 156 CMR Images | | | | | |
| [283] | Accurate Ventricular | UK Biobank | 4848 Subjects | -- | I2-GAN | -- | NA | CC LV=99.91 |
| | Volume Measurements | | | | | | | |
| [284] | Accelerated Multi-Channel CMR | Publicly available abdominal | 28 Subjects | Undersampling | PIC-GAN | Softmax | TensorFlow | Different Results |
| | | Knee | 20 Subjects | | | | | |
| [285] | NA | 2019 MS-CMRSeg | 45 Subjects | -- | STN | -- | NA | Different Results |
| [286] | Reconstruction | Clinical | 58 Subjects | | FCN | -- | PyTorch | $R^2$=95.00% |
| [287] | Reconstruction | Clinical | 35 Subjects | -- | U-Net | -- | NA | ESV=0.1 ml<br>EDV= −0.9 ml |
| [288] | Reduce Motion Artifacts | MICCAI 2017 | 159 Subjects | 2D FFT | RNN | -- | PyTorch | SSIM=88.40<br>PSNR=28.51 |
| | | Cedars | | | | | | |
| [289] | Enhance Spatial Detail | Clinical | 367 Subjects | -- | 2D-CNN | -- | Keras TensorFlow | LV EF=64 |
| [290] | Reduce Scan Time | Clinical | 108 Subjects | NUFFT, IFFT | MD-CNN | -- | PyTorch | SSIM=87<br>MSE=11<br>DM LV=98% |
| [291] | Reconstruction | 3D LGE CMR | 219 Subjects | 3D IFFT | CNN | -- | PyTorch | SSIM=87.6<br>MSE=7.7 |
| [292] | Spatial Resolution of CMR | Clinical | Different Subjects | DA, IFFT | 4DFlowNet | Sigmoid | TensorFlow | Flow Rate= 10.7 |
| | | MICCAI 2012 | | | | | | |
| [293] | Reconstruction | Clinical | 22 Subjects | DA | DL-ESPIRiT | -- | TensorFlow | -- |
| [294] | Reconstruction | MICCAI 2013 | Different Datasets with Subjects | Using IFT and UFT Transform | NISTAD | -- | NA | SSIM=98 |
| [295] | Automatically Compute Cardiac Views | Clinical | 391 Subjects | -- | 3D Extension of the 2D ENet | -- | NA | Different Results |
| [296] | Reconstruction | Clinical | 10 Subjects | ?? | Deep Cascade of CNNs | -- | NA | -- |
| [297] | Reconstruction | Clinical | Different Subjects | ?? | MoDL-STORM | -- | TensorFlow | -- |
| [298] | Produce CMR Images | MICCAI 2017 | 2980 Slices | DA | Proposed Model | -- | NA | Different Results |
| | | Sunnybrook | 714 Slices | | | | | |
| [299] | Artefact Detection | UK Biobank | 3465 CMR Images | ROIs Extraction, DA | 3D Spatio-Temporal CNNs | Softmax | Keras TensorFlow | Acc= 98.20%<br>Prec= 80.90% |
| [300] | Super Resolution CMR | Clinical | 64 Subjects | DA | LSRGAN | -- | NA | -- |
| [301] | Motion Correction in CMR | Clinical | 192 Subjects | ??? | Adversarial Autoencoder Network | -- | TensorFlow | -- |
| [302] | Analysis of MYO Native T1 Mapping Images | Clinical | 665 Subjects | ROIs Extraction, DA | FCN | Softmax | TensorFlow | DM=85.00% |
| [303] | Undersampling Artefact Reduction | Clinical | 19 Subjects | -- | Modified U-Net | -- | NA | -- |
| [304] | LV volumes and function | Clinical | 50 Subjects | -- | Inspired by U-Net | -- | NA | LV ESV=73.1<br>LV SV=78.8<br>LV EF=52.2 |
| [305] | Reconstruction | Clinical | 4 Subjects | -- | MoDL-SToRM | -- | TensorFlow | -- |
| [306] | Reconstruction | Clinical | 178 Subjects | -- | Cascaded CNN Models | -- | NA | Different Results |
| [307] | LV segmentation | Sunnybrook | 15 Subjects | -- | P-GAN, U-Net | -- | TensorFlow | -- |
| [308] | Cardiovascular MR Scans | Clinical | 159 LGE CMR Scans | DA | ScarGAN U-Net | -- | NA | DM:<br>LV End=89.90%<br>LV Epi=90.60% |
| [309] | Reconstruction | MICCAI 2013 | Different Subjects | -- | DA-FWGAN | -- | Python and TensorFlow | Different Results |
| [310] | 4D Semantic CMR Synthesis | MICCAI 2017 | 100 Subjects | -- | SPADE GAN | -- | NA | -- |
| [311] | generating CMR Slices | UK Biobank | 402 Subjects | -- | SPSGAN | -- | NA | -- |
| [312] | Medical synthetic images | Clinical | 292 Subjects | ROIs Extraction | E-GAN and SimGAN | ResNet-50 and Xception | TensorFlow Keras | Acc=88.00% |
| [313] | Congenital heart disease | Clinical | 345 Subjects | DA | PG-GAN | Softmax | TensorFlow | DM LV: 97.80% |
| [314] | Simulation for cardiac fiber structure | Clinical | 246 Subjects | ?? | DCGAN | -- | NA | -- |
| [315] | motion artifact correction | Clinical | 60 Subjects | DA | ResNet | Sigmoid + Convolution | PyTorch | PSNR=31.22 |
| [316] | MYO infarction Classification | Clinical | 73 Subjects | -- | SDAE | SVM | PyTorch | Acc=87.60%<br>Prec=86.20% |
| [317] | Full LV Coverage | UK Biobank | 800000 volumes | DA | Fisher-Discriminative 3D-CNN | Fisher | NA | Prec:91.81% |
| | | Sunnybrook | 1120 volumes | | | | | |
| [318] | Cardiac Indices | MICCAI 2019 | 56 Subjects | -- | U-Net and DenseNet | -- | NA | DM=93.00% |
| [319] | Reconstruction | Clinical | 45 Subjects | 2D IFFT | DeepT1 | FC | NA | -- |
| [320] | LV cavity, RV, MYO at ED and ES | MICCAI 2017 | 100 Subjects | DA | DCNN | -- | PyTorch | -- |
| [321] | CMR Orientation and Segmentation | MyoPS | 45 Subjects | -- | CMRadjustNet | -- | NA | Acc=98.70% |
| | | MICCAI 2017 | 100 Subjects | | | | | |
| [322] | LV Segmentation | Sunnybrook | 45 CMR Images | -- | Registration Network | -- | TensorFlow | DM=93.00% |

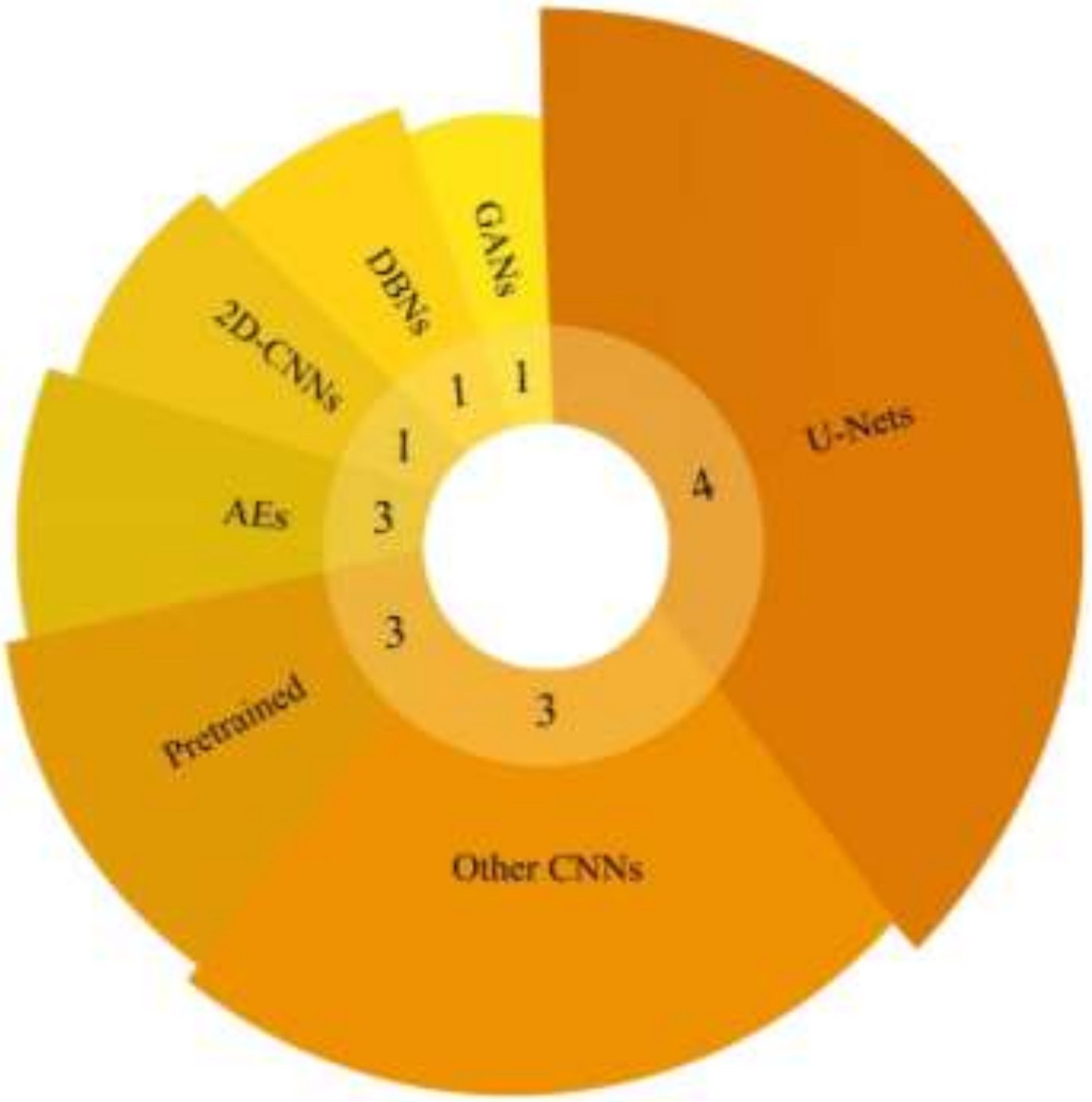

Fig.16. Deep learning conducted for diagnosing CVDs from multimodality data.

## 4.7. Applications of DL for diagnosis of CVDs based on multimodal data

Much research is currently being conducted using multimodality to diagnose various diseases. In clinical research, multimodality has proved successful in diagnosing diseases. To this end, research has been presented in the field of CVDs diagnosis by combining CMR data with other medical imaging methods. In Table (7) and figure (16), the papers on diagnosing CVDs based on multimodalities using DL techniques are reported. As can be seen, to achieve high diagnostic accuracy for various CVDs, researchers have combined the CMR modality with other medical imaging modalities such as CT.

Table 7. Summary of deep learning studies conducted for diagnosing CVDs from multimodality data.

| Ref | Application | Dataset | Modalities | Number of cases | Preprocessing | DNN | Classifier | Toolbox | Performance |
|---|---|---|---|---|---|---|---|---|---|
| [323] | Multi-task Image Segmentation | OASIS project | CMR and CT | Different Subjects | -- | 2D-CNN | Softmax | -- | -- |
| [324] | LV Segmentation | CETUS Challenge 2014 | Echo | 45 Subjects | ROIs Extraction | SAE,GVF-Snake | -- | -- | DM ED=11.20%<br>DM ES=16.00% |
| [325] | Direct LV Estimation | Clinical | 3D Echo | 150 Subjects | -- | CDBN | RF | -- | Different Results |
| [326] | Cardiac substructures segmentation | STACOM 2017 | CT and CMR | 20 MR and 20 CT Images for Training, 40 Test Images | DA | MO-MP-CNN | Softmax | TensorFlow | Sen=83.10%<br>Spec=99.90%<br>Prec=86.80% |
| [327] | Whole Heart Segmentation | MMWHS Challenge | CT and CMR | 20 Contrast-Enhanced CT Scans and 20 CMR | -- | U-Net | -- | Keras, Theano | Different Results |
| [328] | Segmentation | MICCAI 2017 | CT and CMR | 20 CMR and 20 CT | DA | PnP-AdaNet | Softmax | -- | DM=63.90% |
| [329] | Estimating Multitype Cardiac Indices | NA | CT and CMR | 2360 CT and 2900 CMR | | | | | |
| [330] | Anatomically Plausible Segmentation | JSRT, Sunnybrook | X-ray and CMR | 247 X-ray and 45 CMR | DA | Post-DAE | RF | Keras | DM=47.00% |
| [331] | Cardiac Segmentation | MICCAI 2017<br>CAMUS | CMR and Ultrasound | 150 Subjects<br>500 Subjects | -- | cVAE | -- | -- | -- |
| [332] | Segmentation | MM-WHS 2017 | CT, CMR | 20 CMR and 20 CT | -- | GANSA | -- | NA | DM=80.10% |

## 5. Challenges

This section discusses the challenges faced during the CVDs diagnosis from using CMR images and DL techniques. Researchers constantly confront multiple challenges when presenting new approaches for diagnosing CVDs, including datasets, DL models, explainable AI, and hardware resources.

### 5.1. Datasets

Datasets are an essential part of DL-based CADS for detecting CVDs. Previously publicly available datasets of CMR data were introduced in Section 4.1. The available datasets of CMR images suffer from a scarcity of subjects, which hinders researchers using state-of-the-art DL models in CVDs diagnosis. Available datasets with CMR image segmentation applications contain confined subjects. In these datasets, there are limited ground truth images for each class. As a result, researchers face challenges when deploying advanced DL models to precisely segment CMR images. There are many types of heart disease for which early diagnosis is of pivotal significance. However, there are no available datasets of CMR modalities for different types of CVDs, which is another challenge.

### 5.2. Multimodality Dataset

CMR imaging is one of the most important screening methods to diagnose CVDs. In clinical applications, it is challenging for physicians to diagnose some CVDs from CMR images. To this end, physicians take advantage of multimodality imaging to diagnose CVDs. In this procedure, medical specialists exploit CMR images and other imaging techniques such as echo to diagnose CVDs. In [333-334], researchers have indicated that the utilization of multimodality imaging methods to obtain a more accurate diagnosis of CVDs.

In [326-332], to the diagnosis of CVDs, researchers have exploited multimodality datasets. It may be noted that there are multimodal datasets available for the diagnosis of CVDs. However, these datasets have limited subjects and few CVDs. Therefore, restricted access to multimodal datasets with various diseases and a large number of subjects is another associated challenge in the dataset section.

Due to these challenges in this field, until now, researchers have not been able to incorporate advanced DL methods using multimodality imaging to diagnose CVDs. Therefore, providing multimodality imaging datasets based on CMR images with a large number of subjects could facilitate valuable research in the field of CVDs diagnosis. Additionally, the availability of multimodality imaging datasets with a large number of subjects allows researchers to develop state-of-the-art DL methods to aid specialist physicians in diagnosing various types of CVDs.

### 5.3. Limitation CMR Data for Training of DL Models

Researchers have advanced in developing DL models, but there are still many challenges in achieving a real tool for diagnosing CVDs using these approaches. As discussed in the previous sections, many studies have been presented to diagnose CVDs from CMR images using DL techniques. However, achieving real diagnosis software requires the development of DL models based on CMR images. The lack of access to huge CMR datasets for researchers is an important challenge. Some papers have used pre-trained [98-100] or DA [71-73] models to overcome these challenges. Although pre-trained and DA techniques have an array of advantages, there are also challenges associated with them. For example, pre-trained models are trained on ImageNet data [98-100]. Researchers have employed these architectures in many works on medical images such as CMR and have achieved satisfactory results [268]. To enhance the effectiveness of pre-trained models, it is better to train them first on grayscale medical images and then use them to diagnose CVDs. In addition, DA methods play an important role in the generation of synthetic medical data for training DL models [71-73]. GAN models are very popular in synthetic data generation, such as CMR data [307]. Although these models have been largely successful in training the model and preventing overfitting of DL models, they need further development for real-world applications in diagnosing CVDs.

### 5.4. DL Models

This paper reviewed researches on the diagnosis of CVDs from CMR images using DL techniques. This section introduces the challenges associated with DL models in CVDs diagnosis research. Standard CNN models are often utilized in papers on CVD diagnosis based on CMR images. CNN models include segmentation and classification architectures in two dimensions [96-97]. Meanwhile, some papers have exploited 3D-CNN models which need a lot of input data for training and have more complex training compared to 2D-CNN models [96-97]. Besides, 3D-CNNs models require strong hardware resources for training [96-97]. Considering these cases, researchers face various challenges in developing 3D-CNN models to diagnose CVDs from CMR images. The lack of trust by physicians in the results of DL models about CVDs is another challenge. DL models normally yield high evaluation parameters (such as accuracy) in diagnosing CVDs from CMR images. The use of uncertainty techniques and explainable approaches can increase the trust of physicians in DL models to diagnose CVDs.

### 5.5. Explainable AI Models

To date, researchers have applied various DL models to diagnose CVDs. As previously discussed, various segmentation and classification techniques based on DL are employed to diagnose CVDs from CMR images. DL-based segmentation techniques are used to extract CVD-related areas from CMR images. DL-based classification models are also used to diagnose CVDs based on CMR images. One of the challenges of DL models in diagnosing CVDs is the failure to identify areas suspected of CVDs in the early stages and indicate them to physicians. Meanwhile, physicians require AI techniques that can diagnose CVDs in the early stages from CMR images. For this purpose, explainable AI methods have been presented by some researchers, which show disease diagnosis in the early stages using medical images [335-336]. For instance, explainable AI has been presented in research to display the early stages of brain tumors from CMR images [337-338]. Providing explainable AI techniques in conjunction with DL models can help specialists to more accurately diagnose CVDs using CMR images.

### 5.6. Hardware Resources

In this section, hardware resources for the implementation of DL models are presented as another challenge. As discussed above, implementing DL methods requires suitable hardware resources. Although a variety of high-performance hardware is offered for various applications, including the implementation of DL models, their high cost has made it impossible for all researchers to use them in research. For example, detecting CVDs from 3D data requires high hardware resources. Due to the lack of access to appropriate hardware resources, researchers convert 3D data to 2D. Additionally, they employ 2D models to detect different types of CVDs. Because implementing 3D DL models has several challenges, including memory shortage, increasing computational load, hardware cost, etc. [96-97]. Although Google and Amazon provide researchers with computing servers to implement advanced DL models [339-340], these tools are not suitable in real-time applications for detecting CVDs from CMR images.

## 6. Discussion

This study accomplished a comprehensive review of CVDs diagnosis research through CMR images using DL methods. CVDs diagnosis using CMR images based on DL techniques is summarized in Tables (4) to (7). A description of each paper is provided in Tables (4) to (7), including the application, dataset, number of samples, preprocessing methods, DL model, implementation tool, and evaluation parameters. A complete comparison is made between all studies in this field in terms of applications, datasets, DL models, and implementation tools. In addition, the current review paper compared with other related works.

### 6.1. Comparison of our work with other review papers

Review papers on diagnosing CVDs using ML and DL is summarized in Table (2). It can be noted from Table (2) that, some researchers have analyzed ML, DL, and ML-DL papers for the segmentation of different parts of the heart. Further, few researchers have proposed review papers on diagnosing CVDs through ECG signals using DL techniques. In this work, papers on the diagnosis of cardiac diseases from CMR images using DL schemes are reviewed. In our paper, all free and publicly available CMR datasets were first reported and then summarized in Table (2). The papers on segmentation, classification and other applications from CMR images using DL models were reviewed. In similar review papers, CMR papers for different applications have not been reviewed so far. Meanwhile, we have reviewed all the research in this field. Additionally, in our review paper, the challenges of diagnosing CVDs from CMR data using various DL methods are discussed in detail. Furthermore, the main future directions in our study are explained in detail, but they are not fully discussed in other reviews. In Figure (17), our work is compared with other review papers in the diagnosis of CVDs using AI methods.

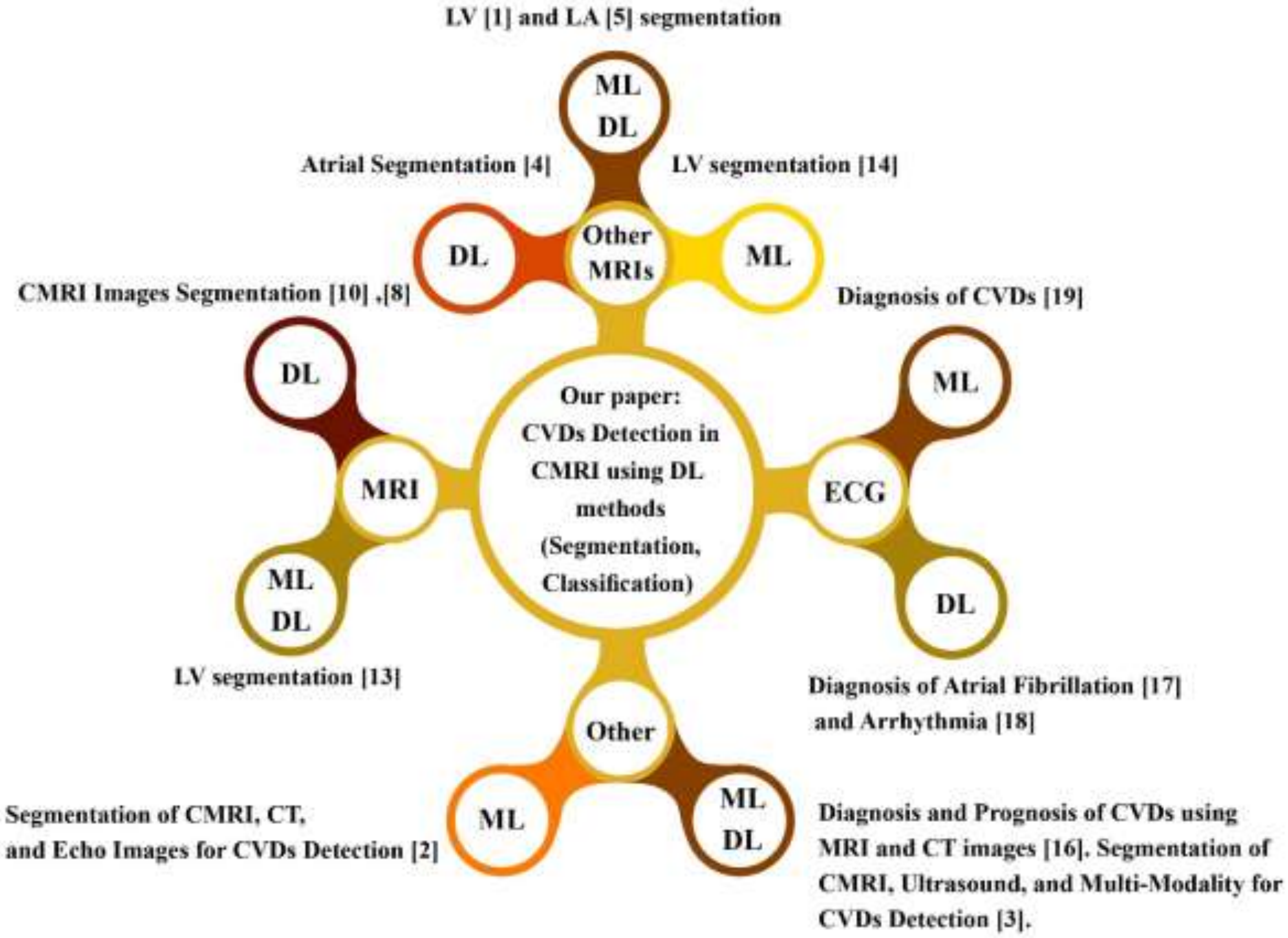

Fig. 17. Comparison of our study with other review papers published on CVDs detection using AI techniques.

## 6.2. Applications

In this section, various applications of DL on CMR data are introduced. The most important section of this paper deals with the presentation of segmentation and classification models based on DL for diagnosing CVDs from CMR images. In Tables (4) and (5), research on CVDs diagnosis using DL-based segmentation and classification techniques were presented, respectively. Other applications of DL models on CMR images are summarized in Table (6). Table (6) indicates few most important applications of DL on CMR images, including reconstruction [296], automatically computing cardiac views [295], and motion artifact correction [301]. Besides, CVDs diagnosis research on the multimodal dataset using DL techniques is also reported in Table (7). The types of DL applications on CMR images are shown in Figure (18). As shown in Figure (18), it can be perceived that the DL-based segmentation methods are the most widely exploited in CVDs diagnosis using CMR images.

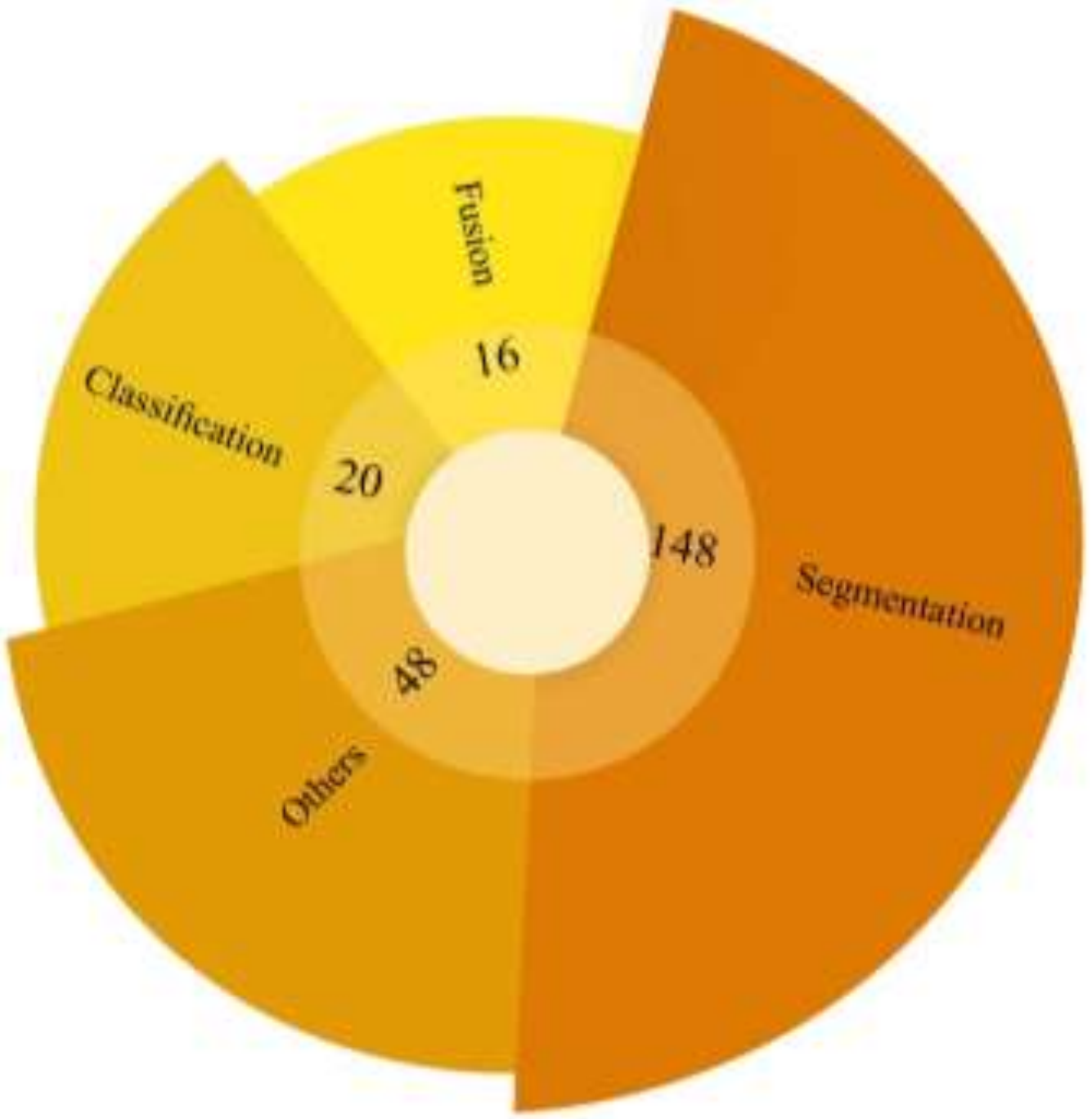

Fig. 18. Deep learning methods used in different applications with CMR images.

**6.3. Datasets**

In section 4.1, a variety of free and publicly available datasets of CMR images were described. It was observed that multiple CMR images datasets have been provided for segmentation and classification applications for diagnosing CVDs. The datasets used in the CVDs research are presented in a part of the Tables (4) to (7). Accordingly, the datasets used for research in this field are shown in Figure (19). Figure (19) illustrates that the researchers focused most on the MICCAI 2017 dataset.

**6.4. Deep learning models**

DL-based applications for CMR images are discussed in this section. An overview of the famous DL models, such as CNN's, RNNs, AEs, GANs, U-Nets, and FCNs, are presented in this review paper. In the following, the types of DL models in CVDs detection research are summarized in Tables (4) to (7). Figure (20) displays the types of DL models used in this field. According to Figure (20), CNNs models are the most commonly used in this field of research. CNN models perform well on medical gray-scale images. As a result, researchers take advantage of CNNs models on CMR images due to their benefits. Additionally, Figure (20) shows that CNN-based segmentation models are highly popular for diagnosing CVDs in CMR images.

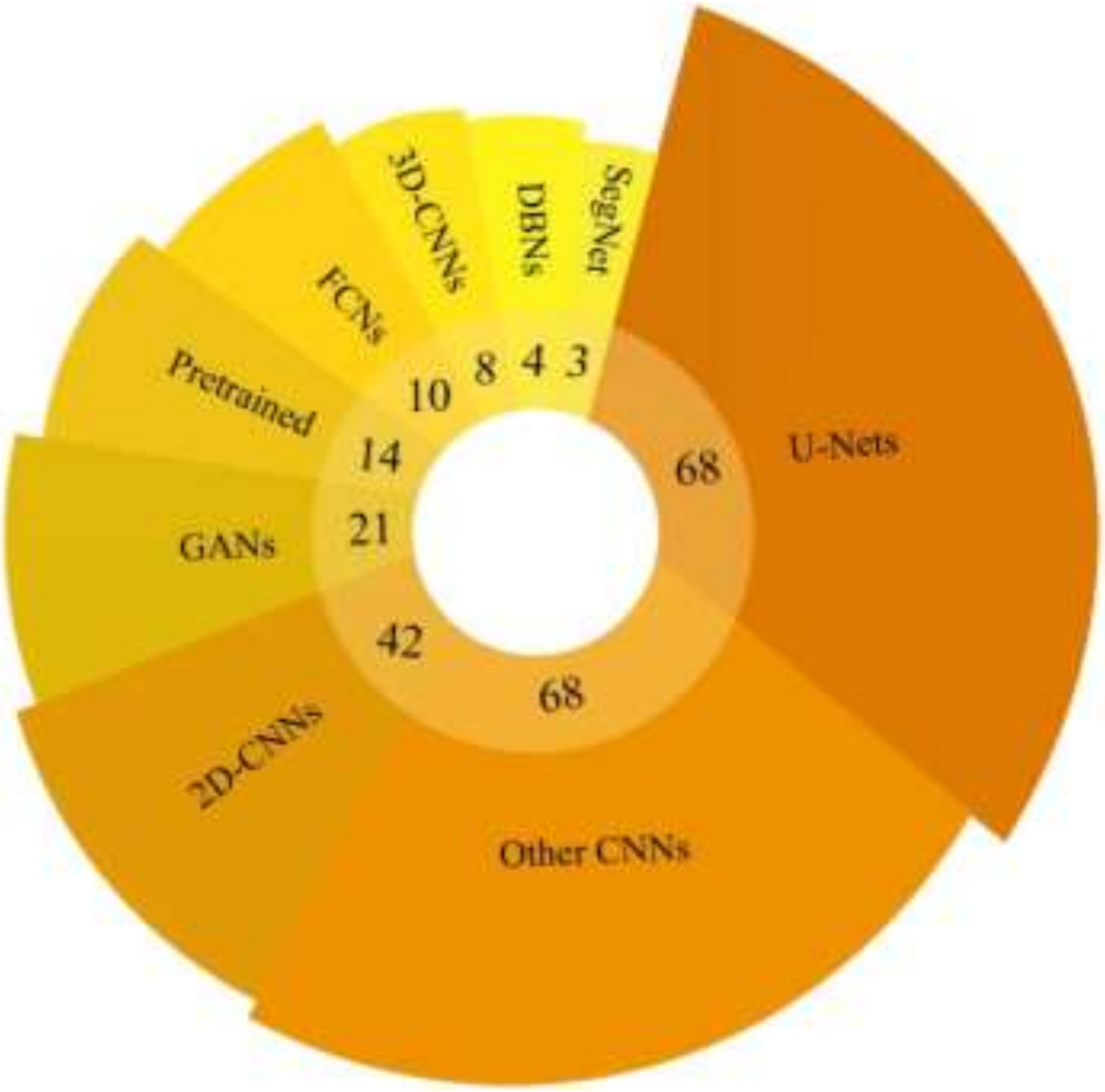

Fig. 19. CMRI dataset used in diagnosis of CVDs using DL methods.

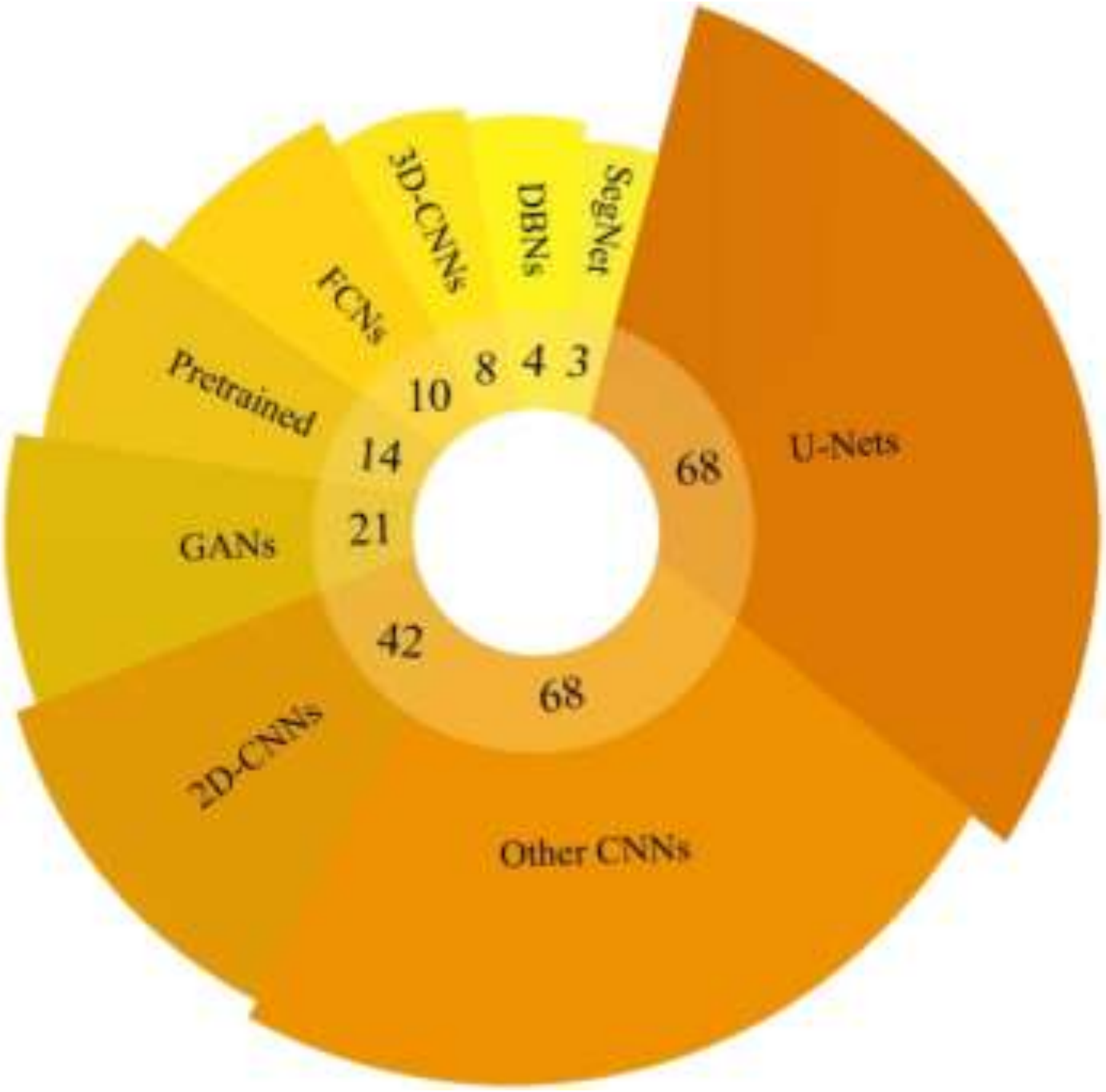

Fig. 20. Deep learning models used for diagnosis of CVDs from CMR images.

### 6.5. Toolboxes

To date, various toolboxes have been developed for implementing DL models. The toolboxes exploited to implement DL models are listed in Tables (4) to (7). TensorFlow, Keras, PyTorch, Theano, and Caffe are the most important toolboxes in DL applications for CMR images. In Figure (21), the types of DL toolboxes are displayed. According to Figure (21), the TensorFlow toolbox has been applied in most researches to implement DL models. Due to its efficiency and simplicity, many researchers employed the Keras Toolbox to implement DL models in CMRI research.

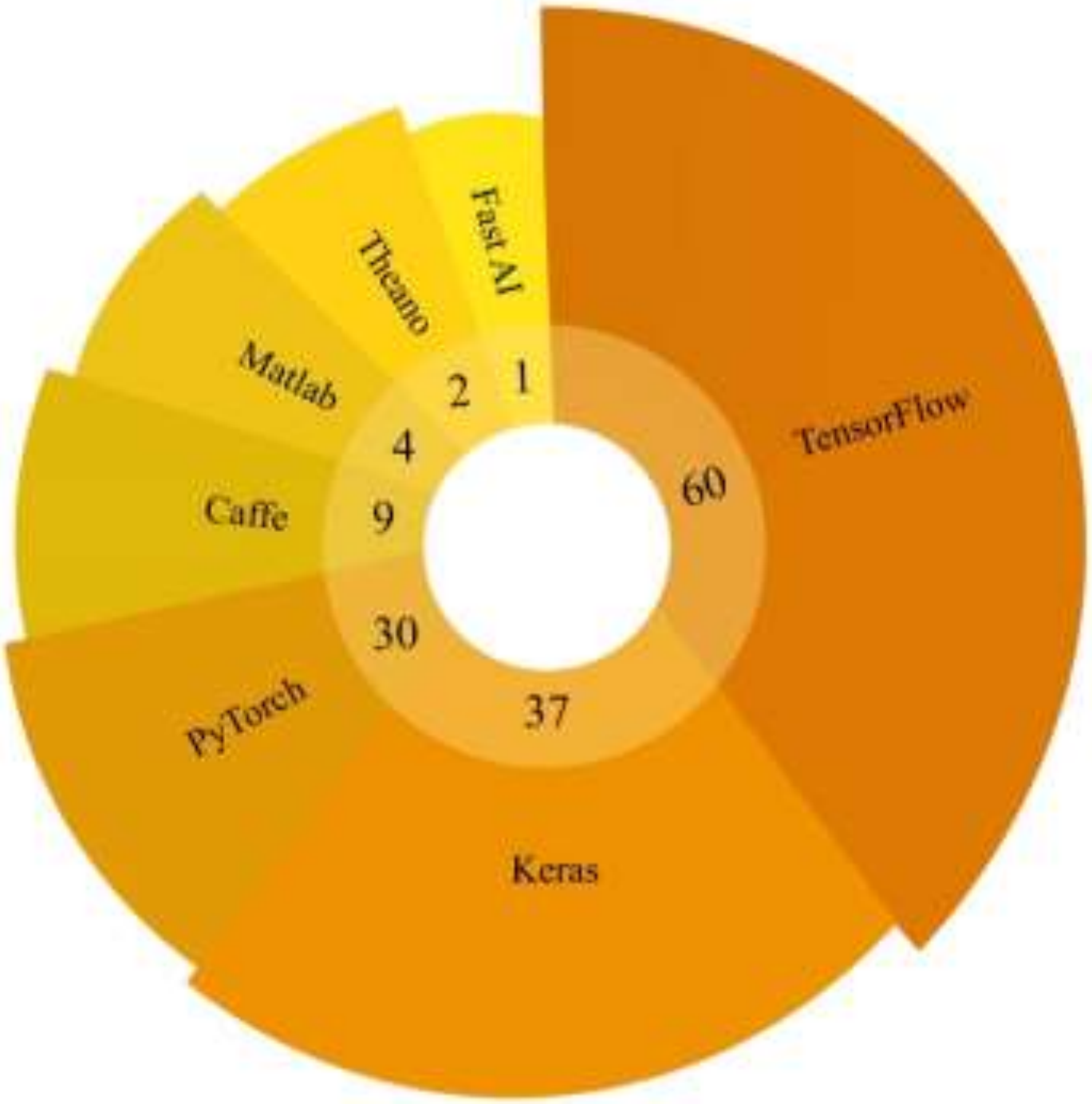

Fig. 21. Deep learning toolboxes used in the diagnosis of CVDs using CMR images.

## 7. Future work

This section outlines potential directions for future research on CVDs diagnosis from CMR images using DL techniques. The challenges related to CVDs detection using CMR data were discussed in the previous section. In this section, suggestions are made to overcome the challenges faced in diagnosing CVDs using DL techniques. Future work in this area includes dataset, multimodal datasets, DL models, explainable AI, and hardware resources. Addressing any challenges in the CVDs diagnosis field can lead to providing real software to assist specialists in the future.

### 7.1. Future works in datasets

This section is dedicated to future work on CVDs datasets containing CMR images. As mentioned earlier, researchers lack access to free datasets with large number of subjects of CMR images for diagnosis of CVDs. As a future work, providing available datasets with many cases from CMR images will bring about important research in CVDs. Additionally, for future work, providing available datasets of CAD, Arrhythmia, cardiomyopathy, CHD, mitral regurgitation, and angina can lead to more applied research.

By providing CMR datasets belonging to large number of subjects, can help to diagnose CVDs using state-of-the-art DL techniques. Also, datasets of CMR images with many subjects have not been made available to researchers to predict various types of CVDs, and this issue can be considered another future work. A discussion of the available datasets for the segmentation of CMR images is presented in Section 4.1. These datasets are provided to diagnose few CVDs. On the other hand, these datasets have limited number of subjects with few CVDs. As another future insight, researchers' access to CMR segmentation datasets with a large number of subjects could yield invaluable research in diagnosing CVDs.

### 7.2. Future Works in Multimodality Datasets

As previously discussed, the diagnosis of diseases by multimodality medical imaging techniques is of particular significance for specialist doctors. A medical imaging method often does not provide all the information necessary to diagnose a disease in full detail. Therefore, multimodality imaging techniques minimize doctors' errors when diagnosing CVDs. In references [333-334] few researchers have taken advantage of multimodality cardiac imaging methods for diagnosing CVDs and obtained promising results. For multimodality imaging research, various datasets available for detecting CVDs are not provided. For future work, researchers' access to datasets of multimodality cardiac imaging methods with many subjects can bring about worthy research to diagnose CVDs using DL techniques. For example, CMR-CT, CMR X-ray, and CMR-Ultrasound datasets have not yet been provided to the diagnosis of myocarditis disease. As a result, the provision of these datasets can yield valuable research in diagnosis of myocarditis disease with DL architectures.

### 7.3. DL Models

This section presents future works for DL methods in diagnosing CVDs using CMR images. As shown in Tables (4) to (7), the researchers used standard DL techniques for CVDs detection. In these studies, researchers have often applied the CNNs, RNNs, AEs, U-Nets, and FCNs models and improved models. Some state-of-the-art DL models include attention mechanism [341-343], transformers [344-345], GANs [74-76], graph CNNs [346-347], and deep reinforcement learning (RL) [348-349]. In the following, details of the state-of-the-art DL methods are discussed.

#### 7.3.1. Deep Attention Mechanism

The attention mechanism is one of the newest areas of DL and has received attention from researchers in diagnosing various diseases [350-351]. These procedures use important input information to predict outputs [341-342]. Attention models are widely diverse, some of the most important of which comprise Attention CNNs [352], attention AEs [353], graph attention [354], and attention RNNs [355]. Researchers can utilize attention mechanism models in CVDs diagnosis using CMR images for future work.

#### 7.3.2. Transformers

Recently, researchers introduced a novel class of DL models called transformers and exploited them in various applications. According to the reference [344], transformer techniques consist of two parts: decoder, and encoder, and use the self-attention architecture [344-345]. ViT is the most significant transformer architecture and has been employed in research for various disease diagnosis using medical data [356-357]. Graph transformers [358], polar transformers [359], and Vit transformers [360] are among the most important models.

#### 7.3.3. Generative Adversarial Networks (GAN)

Another direction that can be investigated in the future is the applications of novel models of GANs in the diagnosis of CVDs. Recent variations of GANs can be used in various possible ways, for example, disentangled representation learning GAN (DR-GAN) and information maximizing GAN (InFoGAN) for representation learning [74-76]. Alternatively, researchers can use GANs for style transfer and unpaired image-to-image translation with Gated-GAN and CycleGAN [74-76]. Moreover, networks such as super-resolution GAN (SRGAN) have introduced methods for improving the quality of data [74-76], which can be used in CADS to help clinicians in their diagnosis.

#### 7.3.4. Graph CNNs

Graph is one of the most popular fields of AI that is of great interest to researchers in ML and DL applications [346-347]. Until a few years ago, graph-based techniques were widely used in ML. However, DL-based graph approaches have recently been introduced, and satisfactory results have been achieved in various aspects [346-347]. DL-based graphs enjoy considerable diversity, some of the most important of which include graph CNNs [346-347], graph RNN [361], etc. In the future, Graph models can be used to diagnose CVDs.

#### 7.3.5. Deep Reinforcement Learning (RL)

This field combines RL and DL and is used to address various problems such as medicine [362-363]. Deep RL models show impressive performance when dealing with large input data and can make optimal problem-solving decisions. Deep Q network (DQN) [364], deep deterministic policy gradient (DDPG) [365], and double DQN (DDQN) [366] are some of the most popular deep RL architectures.

### 7.4. Explainable AI

As mentioned in the previous section, DL models exploit segmentation or classification methods for diagnosing CVDs. However, physicians tend to distrust DL methods to diagnose CVDs using CMR images in the early stages. Because DL models are not efficient at detecting CVDs in the early stages from CMR images. To address this issue, explainable AI methods have been presented, which can be used as a post-processing step in DL-based CADS to the diagnosis of diseases. In future, explainable AI techniques [335-336] can be used to visualize the decisions made by AI by visualizing the abnormality. Hence, it helps to provide confidence to the clinicians in the automatic diagnosis of CVDs from CMR images using DL techniques.

### 7.5. Hardware Resources

In the previous section, lack of access to hardware resources was presented as an important challenge. To deal with this issue, researchers have proposed several techniques for the efficient implementation of DL models. The utilization of quantization and compression techniques for DL networks can be introduced as future work in this field [367-368]. Quantization and compression techniques greatly reduce the demanded computations in DL models [367-368]. This leads to deploying the proposed DL model on a computer, requiring fewer hardware resources. Recently, deep compact-size CNNs techniques have been introduced that do not require more powerful hardware resources to be implemented [369]. The most important models of deep compact-size CNNs are TinyNet [370] and MobileNets models [371].

## 8. Conclusion and findings

CVDs cause an adverse impact on the structure and function of the heart muscle, endangering human health worldwide. CAD, arrhythmia, heart failure, myocarditis, and HCD are the most critical CVDs [1-6]. CVDs are conditions affecting the heart or blood vessels and are usually due to the accumulation of fatty deposits inside the arteries and an increased risk of blood clots [9-11]. Uncontrolled high blood pressure can lead to hardening and thickening of the arteries and narrowing the vessels through which blood flows [14]. According to the Centers for Disease Control and Prevention (CDC), CVDs are the leading cause of death in the United States [15-17].

To date, various screening approaches for CVDs diagnosis have been introduced by specialists. ECG [19], Echo [27], exercise stress test [cite], carotid ultrasound [28], CT-Scan [32], and CMR images [35-36] are among the most significant methods for diagnosis of CVDs. On account of its merits, in recent years, CMR imaging has been recognized as one of the best diagnostic techniques used for CVDs by specialist physicians [35-36].

To examine the structure of the heart, physicians use CMR images in CVDs diagnosis. The advantages of CMR data involve the absence of ionizing radiation, superior soft tissue contrast resolution, and high resolution [35-36]. However, despite the advantages mentioned above, CMR images are affected by different artifacts [36]. Additionally, analyzing CMR data is highly time-consuming and labor-intensive for specialist physicians due to the large number of slides recorded. To alleviate these challenges, extensive research is being conducted on the detection of CVDs on CMR images using DL models. This literature investigated the detection of CVDs from CMR images using DL models.

In the introduction section, a comprehensive discussion of CVDs, diagnostic methods in conjunction with advantages and disadvantages, the importance of DL techniques in CVDs diagnosis, and, ultimately, the structure of the review paper are discussed. This section discusses the merits and demerits of using medical imaging techniques such as CMR data to diagnose CVDs.

In the search strategy section, according to the PRISMA guidelines three levels of analysis were performed to select the papers. Additionally, the inclusion and exclusion criteria for selecting papers on the diagnosis of CVDs were summarized in Table (1).

In section 3, review papers published on detecting CVDs from CMR images were investigated using AI methods. First, each review paper was briefly described, and then their details were summarized in Table (2). At the end of this section, the novelty of our review paper is compared with other published works.

Section 4 introduced DL-based CADS to diagnose CVDs using CMR images. First, CADS steps were presented, including datasets, preprocessing techniques, and DL methods. In the following, the research on the diagnosis of CVDs from CMR data are summarized in four Tables: 1) segmentation, 2) classification, 3) other research, and 4) fusion research. According to this section, it was observed that researchers have carried out the most research in the field of CMR image segmentation to the diagnosis of CVDs.

The most important challenges for the diagnosis of CVDs from CMR images were reported in Section 5. This section discusses the challenges of datasets, multi-modality, DL models, explainable AI, and lack of hardware resources accessibility in diagnosing CVDs. Addressing existing challenges allows researchers to access applications to detect CVDs from CMR images.

The discussion section contains several subsections and important information on CVDs detection research is reported based on Tables (4) to (7). This section includes datasets, types of CVDs, applications, DL models, DL implementation tools, and classification techniques.

Section 7 is dedicated to future work on the detection of CVDs from CMR images using DL methods. Future directions are outlined for datasets, DL models, explainable AI, and hardware resources. A complete explanation is introduced for each of the future tasks, along with few novel approaches. This section allows researchers to exploit new ideas for datasets, DL models, and hardware resources in future research.

In recent years, invaluable research has been conducted to detect CVDs from CMR images using various AI techniques. Research shows that it will be possible to achieve real CVDs detection tools based on DL algorithms in the near future. We feel that, state-of-the-art technologies such as telemedicine [372] and IoMT [373] need to be applied with DL models to diagnose CVDs accurately.

## Appendix A: Performance metrics

Accuracy is the ratio of correctly predicted observations to the total number of observations [88].

$$Acc = \frac{TP + TN}{FP + FN + TP + TN}$$

Sensitivity, or recall, is the ratio of the number of correctly predicted positive observations to the total number of cases with the condition of interest [88].

$$Sen = \frac{TP}{FN + TP}$$

Specificity is the number of correctly predicted negative observations to the total number of negative observations [88].

$$Spec = \frac{TN}{FN + TN}$$

Precision, or positive predictive value, is the number of correctly predicted positive observations to the total number of positive observations [88].

$$Prec = \frac{TP}{TP + FP}$$

F-score is the harmonic mean of precision and recall. F-core is preferred for datasets with imbalanced numbers of cases with and without the condition of interest [88].

$$F - Score = \frac{2\,TP}{2\,TP + FP + FN}$$

The dice coefficient, or Sørensen–Dice index, measures the similarity between two datasets [4-5].

$$Dice = \frac{2\,TP}{2\,TP + FP + FN}$$

Hausdorff distance, or Pompeiu–Hausdorff distance, measures how far two subsets of a metric space are from each other [4-5].

$$HD(A_s, B_s) = max\left\{\max_{a \in A_s} \min_{b \in B_s} d(a, b), \max_{b \in B_s} \min_{a \in A_s} d(b, a)\right\}$$

The Jaccard index, or the Jaccard similarity coefficient, measures the similarity and diversity of sample sets [4-5].

$$JAC(R, G) = \frac{|R \cap G|}{|R \cup G|}$$